\documentclass[twocolumn]{aastex63}
\usepackage{graphicx}
\usepackage{color}
\usepackage{amsmath,amssymb}
\usepackage{gensymb}
\usepackage{times}
\usepackage{nicefrac}
\newcommand{\kms}{km s$^{-1}$}

\begin{document}
\author[0000-0002-6993-0826]{Emily C. Cunningham}
\affiliation{Center for Computational Astrophysics, Flatiron Institute, 162 5th Ave., New York, NY 10010, USA}
\author[0000-0001-7107-1744]{Nicolas Garavito-Camargo}
\affiliation{Steward Observatory, University of Arizona, 933 North Cherry Avenue, Tucson, AZ 85721, USA}
\author[0000-0001-6146-2645]{Alis J. Deason}
\affiliation{Institute for Computational Cosmology, Department of Physics, University of Durham, South Road, Durham DH1 3LE, UK}
\author[0000-0001-6244-6727]{Kathryn V. Johnston}
\affiliation{Department of Astronomy, Columbia University, 550 West 120th Street, New York, NY, 10027, USA}
\affiliation{Center for Computational Astrophysics, Flatiron Institute, 162 5th Ave., New York, NY 10010, USA}
\author[0000-0002-8448-5505]{Denis Erkal}
\affiliation{Department of Physics, University of Surrey, Guildford GU2 7XH, UK}
\author[0000-0003- 3922-7336]{Chervin F. P. Laporte}
\affiliation{Kavli IPMU (WPI), UTIAS, The University of Tokyo, Kashiwa, Chiba 277-8583, Japan}
\author[0000-0003-0715-2173]{Gurtina Besla}
\affiliation{Steward Observatory, University of Arizona, 933 North Cherry Avenue, Tucson, AZ 85721, USA}
\author[0000-0002-0296-3826]{Rodrigo Luger}
\affiliation{Center for Computational Astrophysics, Flatiron Institute, 162 5th Ave., New York, NY 10010, USA}
\author[0000-0003-3939-3297]{Robyn E. Sanderson}
\affiliation{Department of Physics and Astronomy, University of Pennsylvania, 209 S 33rd St., Philadelphia, PA 19104, USA}
\affiliation{Center for Computational Astrophysics, Flatiron Institute, 162 5th Ave., New York, NY 10010, USA}

\title{Quantifying the Stellar Halo's Response to the LMC's Infall with Spherical Harmonics}
\begin{abstract}
The vast majority of the mass in the Milky Way (MW) is in dark matter (DM); we therefore cannot directly observe the MW mass distribution, and have to use tracer populations in order to infer properties of the MW DM halo. However, MW halo tracers do not only feel the gravitational influence of the MW itself. 
Tracers can also be affected by MW satellites; \cite{GaravitoCamargo2019} (hereafter GC19) demonstrate that the Large Magellanic Cloud (LMC) induces a density wake in the MW DM, resulting in large scale kinematic patterns in the MW stellar halo. 
In this work, we use spherical harmonic expansion (SHE) of the velocity fields of simulated stellar halos in an effort to disentangle perturbations on large scales (e.g., due to the LMC itself as well as the LMC-induced DM wake) and small scales (due to substructure). Using the GC19 simulations, we demonstrate how the different terms in the SHE of the stellar velocity field reflect the different wake components, and show that these signatures are a strong function of the LMC mass. An exploration of model halos built from accreted dwarfs \citep{Bullock2005} suggests that stellar debris from massive, recent accretion events can produce much more power in the velocity angular power spectra than the perturbation from the LMC-induced wake. We therefore consider two models for the Sagittarius (Sgr) stream --- the most recent, massive accretion event in the MW apart from the LMC --- and find that the angular power on large scales is generally dominated by the LMC-induced wake, even when Sgr is included. We conclude that SHE of the MW stellar halo velocity field may therefore be a useful tool in quantifying the response of the MW DM halo to the LMC's infall. 
  
\end{abstract}


\section{Introduction}
\label{sec:intro}

A fundamental challenge facing the study of the Milky Way (MW) galaxy is that most of its mass is in dark matter (DM). Because we cannot directly observe the MW's DM halo, we must use tracer populations, such as halo stars, globular clusters, and MW satellites, to study the MW's DM halo indirectly. In particular, the velocity distributions of tracer populations can be used to derive estimates of the MW's mass, utilizing methods derived from \cite{Jeans1915} modeling (e.g., \citealt{Dehnen2006}, \citealt{Watkins2009}, \citealt{Gnedin2010}, \citealt{Deason2012}, \citealt{Eadie2017}, \citealt{Sohn2018}, \citealt{Watkins2019}, \citealt{Wegg2019}). Prior to the era of \textit{Gaia}, underpinning essentially all studies of the global kinematic structure of the MW stellar halo, as well as estimates of the mass using Jeans modeling, are three key assumptions: that the halo is in equilibrium, isotropic, and phase-mixed. In our current era with access to full phase space information for halo tracers and detailed high resolution simulations, we can confront the ways in which these assumptions are violated, and use this information to understand our Galaxy on a deeper level. 

One major source of disequilibrium in the MW is its most massive satellite, the Large Magellanic Cloud (LMC). The classical picture of the LMC is as a relatively low mass ($\sim 10^{10} M_{\odot}$) satellite orbiting the MW on a $T\sim 2$ Gyr orbit (e.g., \citealt{Avner1967}, \citealt{Hunter1969}, \citealt{Murai1980}, \citealt{Lin1982}, \citealt{Lin1995}, \citealt{Bekki2005}, \citealt{Mastropietro2005}, \citealt{Connors2006}). However, proper motion measurements of the LMC using the Hubble Space Telescope (\textit{HST}) revealed that the total velocity of the LMC is much higher than previously measured ($v\sim 320$ \kms; \citealt{Kallivayalil2006}, \citeyear{Kallivayalil2013}). This high velocity, near the escape speed of the MW,  indicates that the LMC is likely on its first infall, based on backward orbital integrations (\citealt{Besla2007}, \citealt{Kallivayalil2013}) and statistical predictions from cosmological simulations (\citealt{Boylan-Kolchin2011}, \citealt{Busha2011}, \citealt{Gonzalez2013}, \citealt{Patel2017}). In addition, there is mounting evidence that the LMC is more massive than previously thought ($\sim 10^{11} M_{\odot}$), including arguments based on models of the Magellanic system (e.g., \citealt{Besla2010}, \citeyear{Besla2012}, \citealt{Pardy2018}); abundance matching (e.g., \citealt{Behroozi2010}, \citealt{Guo2010}, \citealt{Moster2010}, \citeyear{Moster2013}); the measured rotation curve of the LMC \citep{vanderMarel2014}; the presence of satellites around the LMC, including the Small Magellanic Cloud (e.g., \citealt{Kallivayalil2018}, \citealt{Erkal2019a}, \citealt{Pardy2019}, \citealt{Patel2020}); the timing argument (\citealt{Penarrubia2016}); and perturbations in the Orphan Stream (\citealt{Koposov2019}, \citealt{Erkal2019b}). 

While concerns about the LMC's influence on dynamics in the MW were raised as early as \cite{Avner1967}, the revised picture of the LMC as a massive ($\sim 10^{11} M_{\odot}$) satellite approaching the MW for the first time is increasingly worrisome for estimates of the MW gravitational potential that neglect the LMC, as the LMC mass is a significant fraction of the MW halo mass. Several studies of simulations of the LMC's infall demonstrate that a massive LMC invalidates the assumption of an inertial Galactocentric reference frame, as the center-of-mass (COM) can be substantially displaced (by as much as 30 kpc; \citealt{Gomez2015}) from the center of the Galaxy, resulting in net motion of the halo with respect to the MW disk. This net COM motion is predicted to be $\sim$ 40 \kms (GC19, \citealt{Erkal2019b}, \citealt{Petersen2020}); \cite{Gomez2015} find that the net motion could be as high as 75 \kms.
\cite{Erkal2020} show that ignoring the influence of the LMC leads to systematic overestimates (as high as $50\%$) of the MW mass when using equilibrium models. Therefore, when considering the motions of tracer populations that we use to study the MW DM halo, we must account for the influence of the LMC in our models.

In addition to perturbations as a result of the COM motion of the LMC, MW halo tracers are also predicted to be perturbed by the LMC-induced DM wake (\citealt{GaravitoCamargo2019}; hereafter GC19). In the $\Lambda$CDM cosmological paradigm, host halos are predicted to respond to the infall of satellites; this response can be thought of as a gravitational or density wake. One component of this wake arises due to local interactions of particles with the satellite. The satellite transfers kinetic energy to nearby resonant particles, creating an overdensity trailing in its orbit and causing an effective drag force on the satellite (i.e., dynamical friction; e.g., \citealt{Chandrasekhar1943}, \citealt{White1983}, \citealt{Tremaine1984}). GC19 refer to this component of the wake as the \textit{Transient response}, as it is expected to weaken over time. In addition to the Transient response, there is also a global response in the DM halo, resulting in large scale over and under densities in the DM halo (e.g., \citealt{Weinberg1989}) that can potentially even excite structure in the disk (\citealt{Weinberg1998}, \citealt{Weinberg2006} \citealt{Laporte2018a}). GC19 refer to this component of the wake as the \textit{Collective response}. For the benefit of the reader, we have included these definitions in Table \ref{tab:defn}. Using detailed \textit{N}-body simulations, GC19 demonstrated that the density wake induced by the infall of the LMC gives rise to distinct, correlated kinematic patterns in the MW stellar halo.

GC19 explored these wake signatures in the context of two MW halo models, one isotropic halo and one radially varying, radially anisotropic halo. They find that while there are similarities in the wake morphology for the two models, there are also key differences: the Transient response is much stronger for the model with the radially anisotropic halo, whereas the Collective response is stronger for the isotropic halo. Therefore, understanding how the velocity anisotropy $\beta=1-\sigma_T^2/\sigma_R^2$ behaves in the MW is important for the predicted morphology of the LMC-induced DM wake. Until relatively recently, our knowledge of the motions of halo tracers was limited to one component of motion, the line-of-sight (LOS) velocity; given this major observational constraint, it was necessary to make assumptions about the tangential motions of stars, and isotropy ($\beta=0$) was the most common assumption. However, simulations predict that $\beta$ should become increasingly radially biased as a function of radius (see, e.g., \citealt{Rashkov2013}, \citealt{Loebman2018}), and $\beta$ in the solar neighborhood is radially biased ($\beta \sim 0.5-0.7$; \citealt{Smith2009}, \citealt{Bond2010}).

We now have full phase space information for distant halo tracers, from the \textit{Gaia} mission and \textit{HST} proper motion (PM) studies, and can measure $\beta$ outside the solar neighborhood directly. Estimates of $\beta$ outside of the solar neighborhood have generally found radially biased $\beta$, using GCs (\citealt{Watkins2019}, \citealt{Sohn2018}) and halo stars (\citealt{Bird2018}, \citealt{Lancaster2019}, \citealt{Cunningham2019b}).

Detecting the halo response to the LMC-induced DM wake would be an exciting advancement in testing our assumptions about the properties of dark matter, as well as providing key constraints on the potential of the MW and the mass and orbital history of the LMC. However, the GC19 simulations give predictions for the response in the context of smooth MW DM and stellar halos. In reality, the MW stellar halo contains a wealth of substructure that is not yet phase-mixed, in the form of stellar streams (e.g., \citealt{Odenkirchen2001}, \citealt{Newberg2002}, \citealt{Belokurov2006}, \citealt{Grillmair2006}, \citealt{Shipp2018}; also see \citealt{Newberg2016} for a recent review) and stellar clouds (e.g., \citealt{Newberg2002}, \citealt{Rocha-Pinto2004}, \citealt{Juric2008}, \citealt{Li2016}). In addition, using a sample of MW halo main sequence turnoff stars from the HALO7D survey (\citealt{Cunningham2019a}), \cite{Cunningham2019b} observed that the estimated parameters of the velocity ellipsoid (i.e., $\langle v_{\phi} \rangle,\sigma_{\phi}, \sigma_{R}, \sigma_{\theta}$) were different in the different survey fields; these differing estimates could be interpreted as evidence that the halo is not phase-mixed over the survey range ($\langle r \rangle = 23$ kpc). They also showed maps of the halo velocity anisotropy $\beta$ in two halos from the \textit{Latte} suite of FIRE-2 simulations (introduced in \citealt{Wetzel2016}), finding that the anisotropy can vary over the range $[-1,1]$ across the sky. Some of the variation in the $\beta$ estimates appeared to correlate with stellar overdensities in the halos, indicating that galactic substructure is at least in part responsible for the different velocity distributions. While some substructure in the halo can be clearly identified as overdensities in phase-space and removed from analysis, the presence of velocity substructure in the halo could complicate attempts to detect signatures of the LMC-induced DM wake. For example, \cite{Belokurov2019} recently argued that the Pisces Overdensity (\citealt{Sesar2007}, \citealt{Watkins2009}, \citealt{Nie2015}) might be stars in the wake trailing the LMC in its orbit, because of their net negative radial velocities. However, it remains difficult to conclusively argue this scenario given that these stars could also be in Galactic substructure (or, perhaps, stars that are in substructure and have been perturbed by the DM wake).

In summary, there is substantial observational evidence that MW stellar halo is in disequilibrium, on average radially anisotropic, and rife with unphase-mixed substructure, in clear violation of the three central assumptions of equilibrium models. The velocity field of the halo contains information about the potential of the MW, the dwarf galaxies that were consumed as the MW assembled its mass, and the properties of its current most massive perturber, the LMC. However, separating out the different origins of the features in the MW halo velocity field remains a formidable challenge.
One way forward is to consider the spatial scale of perturbations: we expect substructure to cause velocity variation on relatively small spatial scales, as opposed to the large scale perturbations from the LMC-induced DM wake. Therefore, to disentangle these effects, we seek a quantitative description of the kinematic structure of the halo that incorporates variation on different spatial scales. Spherical harmonic expansion is a natural tool to address this problem. 

While ideally we would embark on a full basis function expansion (BFE) of the phase space structure of the halo, in this work, we focus on the spherical harmonic expansion of the three components of motion in spherical coordinates, over different distance ranges in the halo (as a complement to this work, Garavito-Camargo et al. 2020, in prep, will present full BFEs of the spatial distributions for these simulations). GC19 explored the density structure and the properties of the velocity dispersions in addition to the mean velocities; we choose to focus on the mean velocities here because of the challenges of estimating densities (i.e., deeply understanding completeness and survey selection functions) and the fact that estimates of the mean of a distribution require fewer tracers than dispersion estimates.

This paper is organized as follows. In Section \ref{sec:sh101}, we present a brief overview of spherical harmonic expansion and define the notation used in the rest of the paper. We then show the results of using spherical harmonic expansion on the velocity fields from the GC19 simulations of the LMC's infall into the MW in Section \ref{sec:lmc_only}. We demonstrate how the different components of the wake are described in terms of spherical harmonics. In Section \ref{sec:bj05}, we investigate how Galactic substructure might complicate our ability to measure perturbations to the velocity field as a result of the LMC-induced DM wake, by studying the \cite{Bullock2005} purely accreted stellar halos. In Section \ref{sec:sgr}, we use two models of the Sagittarius stream to estimate how the MW's most massive stream might influence the angular power spectrum of the MW halo velocity field and interfere with signatures from the wake. We summarize our conclusions in Section \ref{sec:concl}. 

\begin{deluxetable*}{cl}

\tablecaption{Useful Definitions }
\tablenum{1}
\label{tab:defn}
\tablehead{& \textit{LMC-Induced Wake Components from GC19}} 

\startdata
Transient Response &  The overdensity trailing the LMC in its orbit, that arises due to local scattering. This component can be\\
& thought of in terms of classical dynamical friction. \\ 
Collective Response &  Refers both to the overdensity in the north (which arises due to particles in resonance with the LMC's orbit) \\ 
   &  and the motion of the MW disk with respect to the new MW-LMC barycenter. The global response of the \\
   & MW halo to the LMC's infall.\\ 
  \hline \hline
  & \textit{Spherical Harmonics} \\
  \hline
  $\theta$ & Colatitudinal angle, measured downward from the $z-$axis; $\cos \theta = z/R $\\
   $\phi $ & Azimuthal angle, angle in the $x-y$ plane measured from the x-axis; $\tan \phi =y/x $\\
      $\ell$         &Order of Spherical Harmonic $Y_{\ell m}$ \\
    $m$ & Degree of Spherical Harmonic $Y_{\ell m}$ \\
    $a_{\ell m}$ & Spherical harmonic coefficient for mode $\ell m$ \\
    $\varphi_{\ell m}$ & Phase of spherical harmonic coefficient $a_{\ell m}$ \\
    $C_{\ell}$ & Average power in mode of order $\ell$; total power is $(2 \ell+1) \times C_{\ell}$\\
    Zonal Spherical Harmonics & Spherical harmonics with $\ell=0$; rotational symmetry about $z-$axis\\
    Sectorial Spherical Harmonics & Spherical harmonics with $\ell=|m|$ \\
\enddata
\tablecomments{We use the function \texttt{arctan2} in NumPy \citep{oliphant2006guide} to compute azimuthal angle $\phi$ and phase $\varphi$, such that both angles take values over the range $[-\pi, \pi]$.}
\end{deluxetable*}

\section{Spherical Harmonics}
\label{sec:sh101}

\begin{figure*}
    \centering
    \includegraphics[width=\textwidth]{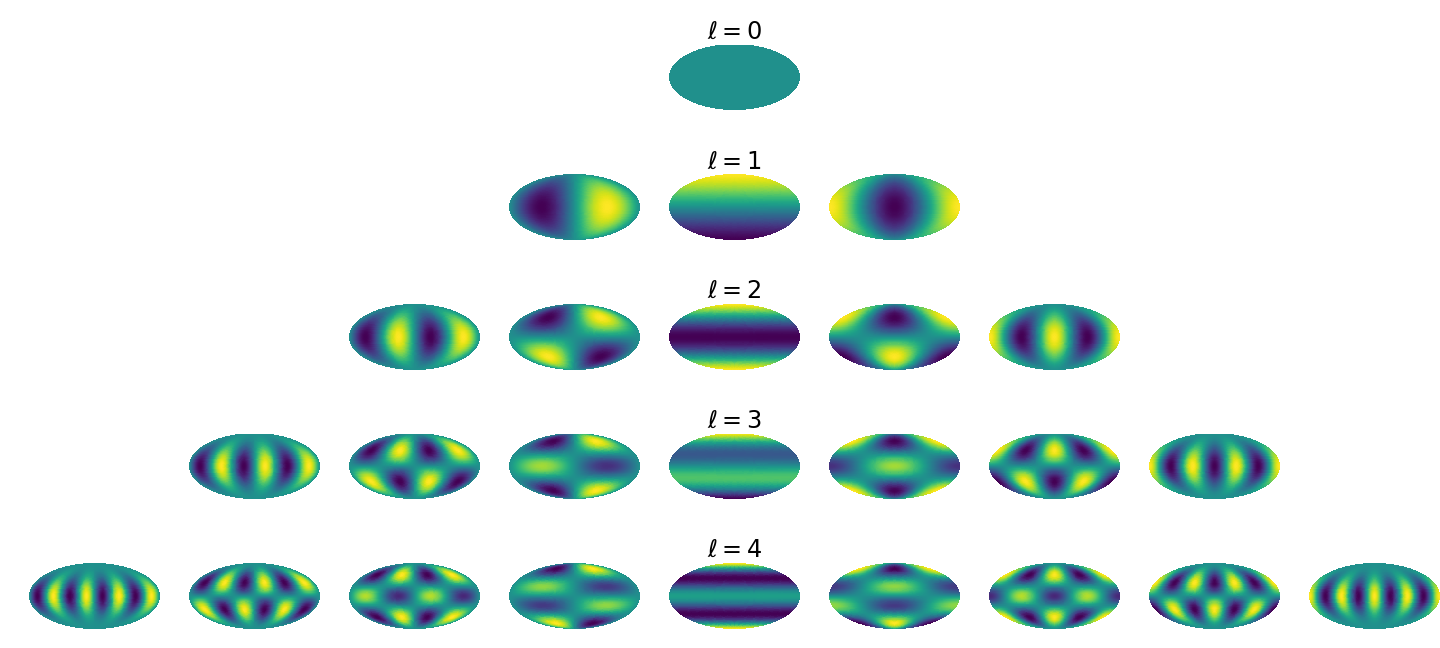}
    \caption{The real spherical harmonics, plotted in Mollweide projection, evaluated from $\ell=0$ to $\ell=4$, for each $-\ell \leq m \leq l$. In this projection, the $z-$axis is oriented upwards, with colatitudinal angle $\theta=0$ at the north pole and $\theta=\pi$ at the south pole. The azimuthal angle runs from $[\pi, -\pi]$ from left to right. The zonal spherical harmonics ($m=0$), which have rotational symmetry about the $z-$axis, are plotted in the central column. The sectorial harmonics ($l=|m|$) are shown in the outermost panels of each row. The only difference between modes with $\pm m$ is a phase shift of 90 degrees.}
    \label{fig:plots}
\end{figure*}

We seek to describe the variation on different spatial scales in halo velocity fields by using spherical harmonic expansion. In this section, we define the notation we use throughout the paper for spherical harmonics. As a reference, we have also included many of these definitions in Table \ref{tab:defn}. Laplace's spherical harmonics of order $\ell$ and degree $m$ are defined as:

\begin{equation}
    Y_{\ell}^{m}(\theta,\phi) = \sqrt{\frac{2 \ell+1}{4 \pi} \frac{(\ell-m)!}{(l+m)!}} P_{\ell}^{m}(\cos \theta) e^{i m \phi},
\end{equation}
where $\theta$ is the colatitudinal angle (i.e., the polar angle measured downward from the north pole) and $\phi$ is the azimuthal angle (i.e., the angle in the $x-y$ plane measured from the $x-$axis), and $P_{\ell}^{m}$ are the associated Legendre polynomials. 

Spherical harmonics comprise an orthogonal basis for any function $f(\theta, \phi)$ defined on the surface of the sphere:

\begin{equation}
    f(\theta,\phi) = \sum_{\ell=0}^{\infty} \sum_{m=-\ell}^{m=\ell} a_{\ell m} Y_{\ell}^{m} (\theta, \phi),
\end{equation}
where $a_{\ell m}$ are the spherical harmonic coefficients are given by:

\begin{equation}
    a_{\ell m}=\int_{\Omega} f(\theta, \phi) Y_{\ell}^{m*}(\theta, \phi) \mathrm{d} \Omega. 
\end{equation}

When the spherical harmonics are complex, the coefficients are also complex; we define the phase $\varphi_{\ell m}$ of a spherical harmonic coefficient as:

\begin{equation}
    \varphi_{\ell m} = \tan^{-1}\left( \frac{\mathrm{Im}[a_{\ell m}]}{\mathrm{Re}[a_{\ell m}]}\right),
\end{equation}
where we use the function \texttt{arctan2} implemented in NumPy (\citealt{oliphant2006guide}) to compute the inverse tangent, taking into account the quadrant in which $a_{\ell m}$ lies in the complex plane.

The angular power spectrum $C_{\ell}$ can be computed from the spherical harmonic coefficients $a_{\ell m}$:

\begin{equation}
    C_{\ell}=\frac{1}{2 \ell +1} \sum_{m} |a_{\ell m}|^2.
\end{equation}
The total power in a given order $\ell$ is thus $(2 \ell +1)\times C_{\ell}$, as there are $(2 \ell +1)$ total $m$ values for a given $\ell$ value. Therefore, in this paper, power spectra will always have the quantity $(2 \ell +1) \times C_{\ell}$ (in units of (\kms)$^2$), as we are expanding the velocity field) plotted on the y-axis.

In this work we use the Python package \texttt{healpy} (\citealt{Zonca2019})\footnote{https://healpy.readthedocs.io/en/latest/}, based on the Healpix scheme \citep{Gorski05}, to perform all of our analysis relating to spherical harmonics. All maps are made using the \texttt{healpy} plotting routine \texttt{mollview}; power spectra and spherical harmonic coefficients are computed using the function \texttt{anafast}; and synthetic maps are generated using \texttt{synfast}.

While \texttt{healpy} works with the spherical harmonics in complex form, the real spherical harmonics are defined as:

\begin{equation}
    Y_{\ell m}(\theta,\phi) = 
    \begin{cases}
P_{\ell m}(\cos{\theta}) \cos(m \phi) & m\geq 0 \\
P_{\ell |m|}(\cos{\theta}) \sin(|m| \phi) & m<0 \\
\end{cases}.
\end{equation}

For illustrative purposes, we show the real spherical harmonics from $\ell=0$ to $\ell=4$ in Figure \ref{fig:plots}. For a given $Y_{\ell m}$, the degree $m$ corresponds to the number of waves along a line of constant latitude. The order $\ell$, in conjunction with degree $m$, determines how many times zero is crossed along a line of constant longitude: there are $\ell - |m|$ zero crossings along a meridian. In the case of $\ell=|m|$, there are no zero crossings along the meridian (outer column in each row of Figure \ref{fig:plots}), and a total of $\ell$ complete waves along the equator; these modes are referred to as the sectorial spherical harmonics. When $m=0$, there are $\ell$ zero crossings along the meridian (central column of Figure \ref{fig:plots}), and no change in amplitude with longitude; these are known as the zonal spherical harmonics, and have symmetry about the $z-$axis. Modes with $m<0$ are of sine type, while modes with $m>0$ are of cosine type; as demonstrated by Figure \ref{fig:plots}, for the real spherical harmonics, changing the sign of $m$ results in a 90$\degree$ rotation about the $z-$axis.

Spherical harmonic expansion and angular power spectra have many applications in astrophysics, most famously in studies of the Cosmic Microwave Background (e.g., \citealt{Planck2018b}, \citealt{Planck2019b}, and references therein). While spherical harmonics are commonly used to describe the angular dependence in full basis function expansions of the potential of dark matter halos (e.g., \citealt{Hernquist1992}; \citealt{Weinberg1996}, \citeyear{Weinberg1999}; \citealt{Lowing2011}), they have not generally been used to describe velocity fields. In the following sections, we discuss spherical harmonic expansion of the velocity fields of several different types of simulations. 

\section{The LMC-Induced Dark Matter Wake}
\label{sec:lmc_only}

In this section, we perform spherical harmonic expansion of the velocity fields on the high resolution, \textit{N}-body simulations of the LMC's infall into the MW from GC19. The kinematic patterns (for mean velocities in all components as well as the densities and velocity dispersions) are discussed in detail in GC19. Here, we discuss how spherical harmonic expansion of the mean velocities can be used to characterize the MW halo response to the LMC's infall. 

This section is organized as follows. We summarize key GC19 simulation properties in Section \ref{sec:gc19_sims}. In Section \ref{sec:fid_sim}, we discuss the spherical harmonic expansion of the velocity maps at 45 kpc in detail. In Section \ref{sec:lmc_d_evolve}, we discuss the radial evolution of the power spectrum, for both the isotropic and radially anisotropic MW models. The dependence of the power spectra on the LMC mass is explored in Section \ref{sec:lmc_mass}. 

\subsection{GC19 Simulation Details}
\label{sec:gc19_sims}

For the full description of the numerical methods employed in these simulations, we refer the reader to Section 3 of GC19. However, we summarize some of the key details here.

The GC19 \textit{N}-body simulations were carried out with Tree Smoothed Particle Hydrodynamics code \texttt{Gadget-3} (\citealt{Springel2008}), with initial conditions specified by the publicly available code \texttt{GalIC} (\citealt{Yurin2014}). The MW model has a virial mass of $M_{\rm MW,vir}=1.2 \times 10^{12} M_{\odot}$, with a DM halo represented by a Hernquist profile with particle masses $m_p=1.57 \times 10^4 M_{\odot}$. The simulations include a disk and bulge component as well, in order to create a realistic potential in the inner halo.  GC19 presents results for both an isotropic halo as well as a halo with radially biased velocity anisotropy. In this work, we focus primarily on the radially anisotropic halo, given that simulations and observations agree that the MW halo should be radially biased (see, e.g., \citealt{Loebman2018}, \citealt{Bird2018}, \citealt{Cunningham2019b}). However, we do discuss the isotropic MW model in Section \ref{sec:lmc_d_evolve}.

For the LMC, they construct four models, with virial masses (prior to infall) of $M_{\rm LMC,vir}=0.8, 1.0, 1.8, 2.5 \times 10^{11} M_{\odot}$. They focus on their fiducial model with $M_{\rm LMC,vir}=1.8 \times 10^{11} M_{\odot}$, which is consistent with LMC mass estimates from abundance matching as well as a first infall scenario (see Section \ref{sec:intro}). 

When considering these simulations, it is important to keep in mind that only the DM is simulated in time, with the stellar halo constructed in post-processing using a weighting scheme (as in \citealt{Laporte2013a}, \citeyear{Laporte2013b}; a generalized version of the scheme used in \citealt{Bullock2005}). The stellar halo is constructed to be in equilibrium with the DM halo, given a specified stellar density and velocity dispersion profile. While in GC19 they construct two stellar halos (one using the K-Giant density profile measured in \citealt{Xue2015}, and one using the density profile as measured from RR Lyrae in \citealt{Hernitschek2018}), for the purposes of this work, we consider only the stellar halo constructed with the \cite{Xue2015} density profile.

GC19 identify two main components of the wake: the Transient and Collective responses. As discussed in the Section \ref{sec:intro}, the Transient response refers to the DM overdensity trailing the LMC in its orbit, corresponding to  the classical \cite{Chandrasekhar1943} wake. The Collective response refers to the global response of the halo to the LMC's infall (\citealt{Weinberg1989}), which results in an extended overdensity in the north, as well as the motion of the MW about the new MW-LMC barycenter. As we refer to these two components of the wake frequently throughout the remainder of the paper, we have included these definitions in Table \ref{tab:defn} as a reference for the reader.  

\subsection{The Velocity Field Near $R_{\rm LMC}$}
\label{sec:fid_sim}

\begin{figure*}
\centering
    \includegraphics[width=0.32\textwidth]{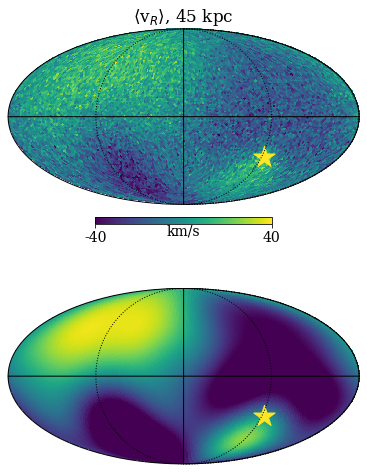}
    \includegraphics[width=0.32\textwidth]{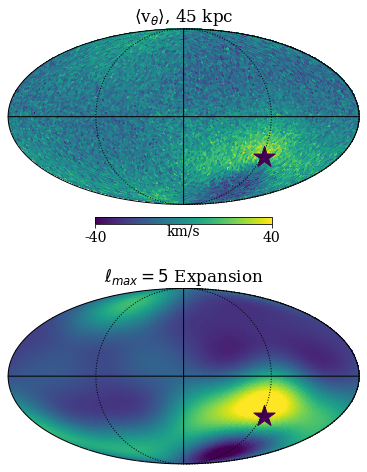}
    \includegraphics[width=0.32\textwidth]{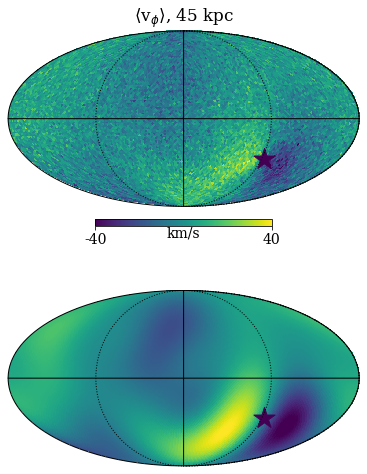}
    \includegraphics[width=\textwidth]{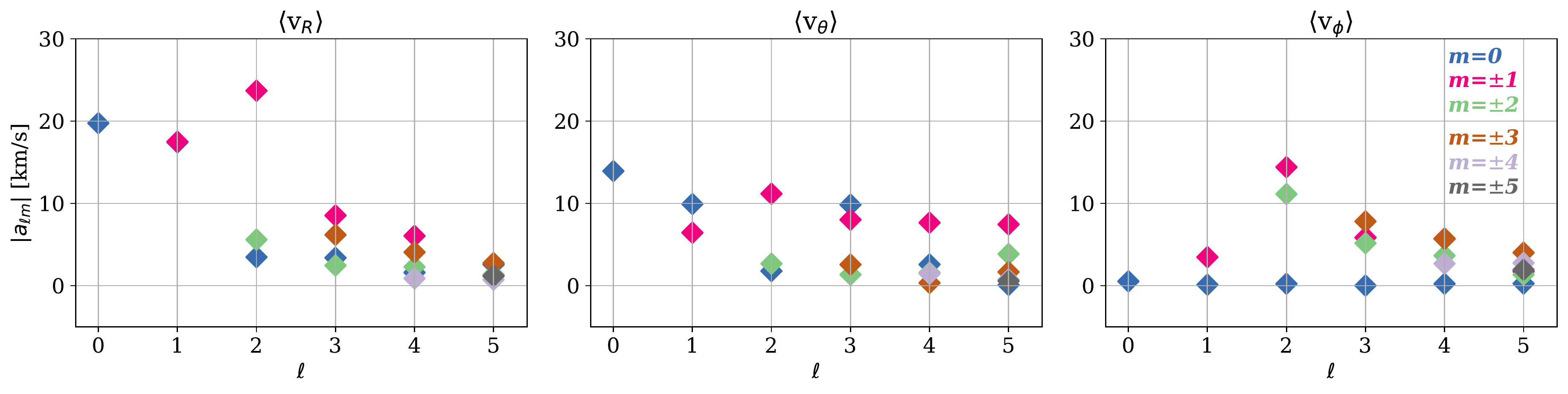}
    \caption{Top panels: average velocity maps, computed in a 5 kpc shell centered on $R=45$ kpc, from the GC19 fiducial simulation ($M_{\rm LMC}=1.8 \times 10^{11} M_{\odot}$) with the radially anisotropic MW model. The average radial velocity map $\langle v_{R} \rangle$ is shown on the left; average polar velocity $\langle v_{\theta} \rangle $ is in the middle panel; and average azimuthal velocity $\langle v_{\phi} \rangle$ is shown on the right. The angular position of the LMC (located at $R=50$ kpc) is indicated by the star. The star is color coded to indicate the sign of the LMC's velocity in each component of motion; the magnitude of the LMC's velocity in all components is greater than the range shown by colorobar ($(v_{R},v_{\theta},v_{\phi})=(99,-345, -46)$ \kms). All velocities and positions are computed with respect to the Galactic center. Middle panel: the $\ell_{\rm max}=5$ spherical harmonic expansion of these maps. A low order spherical harmonic expansion expansion effectively captures the salient features in these velocity maps. Bottom panels: magnitudes of the spherical harmonic coefficients, color coded by degree $m$. For the radial velocity, the dominant mode is $\ell=2, m=\pm 1$; this mode captures the net outward motions of particles in the Transient response (near the LMC) as well as the outward motions of particles in the Collective response (in the north). In $v_{\theta}$, the monopole term is dominant ($\ell=m=0$), reflecting the net motion of the halo with respect to the MW disk, as a result of the new MW-LMC barycenter. In $v_{\phi}$, the $\ell=2, m=\pm 1$ mode dominates; this sectorial mode captures the converging motions of particles trapped in the Transient response, moving towards the orbit of the LMC.}
    \label{fig:45kpc}
\end{figure*}

We first discuss the velocity field from the GC19 fiducial LMC model ($M_{\rm LMC}=1.8 \times 10^{11} M_{\odot}$) with the radially anisotropic MW model (referred to as ``Model 2" in GC19) at 45 kpc (in the Galactocentric frame), very near the present-day position of the LMC ($R_{\rm LMC}=50$ kpc). We expect the velocity maps over this radial range to be sensitive to the Transient response, given that the LMC passed through very recently, and the Transient response arises due to local scattering. The mean velocity maps in three components of spherical motion (with respect to the Galactic center) are shown in the top panels of Figure \ref{fig:45kpc}. 

At $z=0$ in these simulations, the LMC is located at 50 kpc, and its angular position indicated by the star symbol in the top panels of Figure \ref{fig:45kpc}. The color of the star symbol indicates the sign of the motion of the LMC in each component of motion: $(v_{R},v_{\theta},v_{\phi})_{\rm LMC}=(99,-345, -46)$ \kms. At 45 kpc, the motions of the stars in the Transient Response near the LMC's position trace the COM motion of the LMC. The net motion outwards in $v_R$ and the converging motions $v_{\theta}$ and $v_{\phi}$ near the LMC's position result from stars in the transient response, accelerating towards the LMC. In addition, stars at 45 kpc are being accelerated towards the overdensity in the North (i.e., the Collective response): this is reflected in the large areas in the north of net positive $v_R$ and net negative $v_{\theta}$.

The middle panels of Figure \ref{fig:45kpc} show the $\ell_{\rm max}=5$ expansion of each component of velocity. Because the kinematic variation induced by the LMC occurs on large scales, the dominant features in the velocity maps are effectively captured by a low-order spherical harmonic expansion. The lower panels of Figure \ref{fig:45kpc} show the magnitudes of the spherical harmonic expansion coefficients ($|a_{\ell m}|$), color coded by $m$ value. In the radial velocity map (left panels), the dominant term is the $\ell=2, m=\pm 1$ mode; this mode captures the outward radial motion in the upper left quadrant and the lower right quadrant (near the LMC), as well as the inward radial motion in the lower left quadrant and upper right quadrant. In the $v_{\theta}$ maps, the monopole term ($\ell=m=0$) is dominant, reflecting the net upwards motion of the halo with respect to the disk. In the $v_{\phi}$ maps, the $\ell=2, m=\pm 2$ mode dominates; the sectorial $\ell=2$ mode captures the converging motions of stars in the Transient Response near the location of the LMC.\footnote{It is worth noting that while the total power in order $\ell$ is invariant under rotation, the amount of power in a given $\pm m$ value is only invariant under rotations about the $z-$axis. Therefore, our choice to orient these simulations in Galactocentric coordinates aligned with the disk is important to keep in mind, and the dominant $m$ values will be very sensitive to the orbital history of the LMC.} The fact that there is no power at $\ell=0$ in $v_{\phi}$ is indicative of the fact that the GC19 MW models have no net rotation. If the MW does have any net rotation (which has been observed, but not with high statistical significance; \citealt{Deason2017}), this would result in power at $\ell=0$ in $v_{\phi}$, which would not interfere with the predicted wake signature.

We note that we have not included error bars on the spherical harmonic coefficients in Figure \ref{fig:45kpc}, nor do we include error bars on our estimates of the power spectra in subsequent figures. This is because the simulations are very high resolution, so the statistical errors on the coefficients are very small; the dominant sources of uncertainty here are in the models, not in the noise in the simulations.

\subsection{Evolution with Distance}
\label{sec:lmc_d_evolve}

\begin{figure*}
    \centering
    \includegraphics[width=0.33\textwidth]{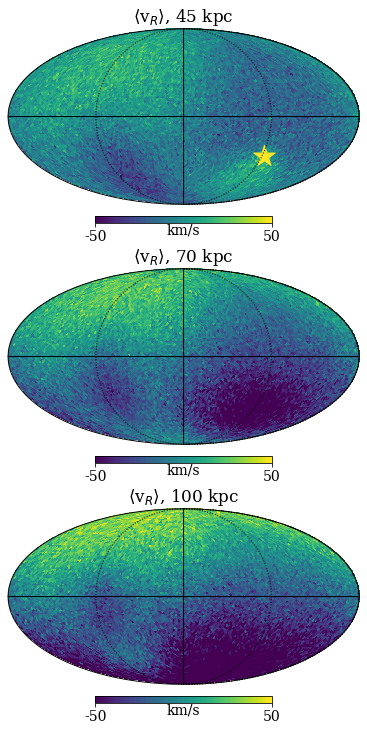}
    \includegraphics[width=0.65\textwidth]{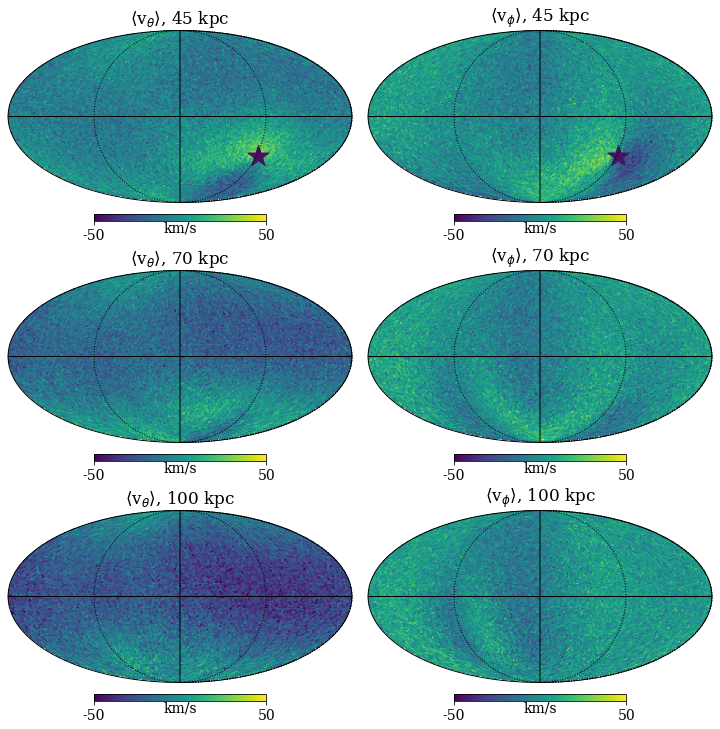}
    \caption{Mean velocity maps in the three components of motion in spherical coordinates ($v_R, v_{\theta}, v_{\phi}$) for the fiducial GC19 simulation with the radially anisotropic MW model. Mean velocity maps are computed in 5 kpc shells. Top panels show the velocity maps at 45 kpc; middle panels show the maps at 70 kpc, and the lower panels show the maps at 100 kpc. As in Figure \ref{fig:45kpc}, the angular position of the LMC is indicated by the star in the top panels, color coded by the sign of the LMC's velocity in each component of motion.}
    \label{fig:maps_3d_v}
\end{figure*}

Figure \ref{fig:maps_3d_v} shows the velocity maps in the three components of spherical motion for this simulation at 45 kpc, 70 kpc, and 100 kpc, in the Galactocentric frame. Radial velocity maps are shown in the left panels, polar motion $v_{\theta}$ is plotted in the middle panels, and azimuthal motion is shown in the right panels. As a result of the Collective Response, there is a radial velocity dipole that increases in strength as a function of distance out into the halo. In addition, the net motion $v_{\theta}$ becomes increasingly negative at larger Galactocentric radii. The behavior in $v_{\theta}$ and $v_{R}$ can also be represented in terms of $v_z$: in these simulations, while the net motion in the plane of the disk is fairly stable ($\langle v_x \rangle = \langle v_y \rangle \sim 0$ \kms ~over all radii), there is net upwards motion in the halo ($\langle v_z \rangle >0$), which increases as a function of distance from the MW's disk. \cite{Erkal2020} show that the MW globular clusters and dwarf satellites also show net motion in $v_z$ (and no net motion in $v_x, v_y$); however, they note the caveat that these tracers may not be phase mixed in the MW potential. We note that the velocity shifts in these simulations result from two sources: the DM overdensity in the north (which gets stronger with distance, because the LMC spends more time at larger radii) as well as the net acceleration of the MW disk towards the LMC (e.g., \citealt{Petersen2020}). Disentangling the relative contributions to the overall velocity shift from these two sources is beyond the scope of this work.

\begin{figure*}
    \centering
    \includegraphics[width=\textwidth]{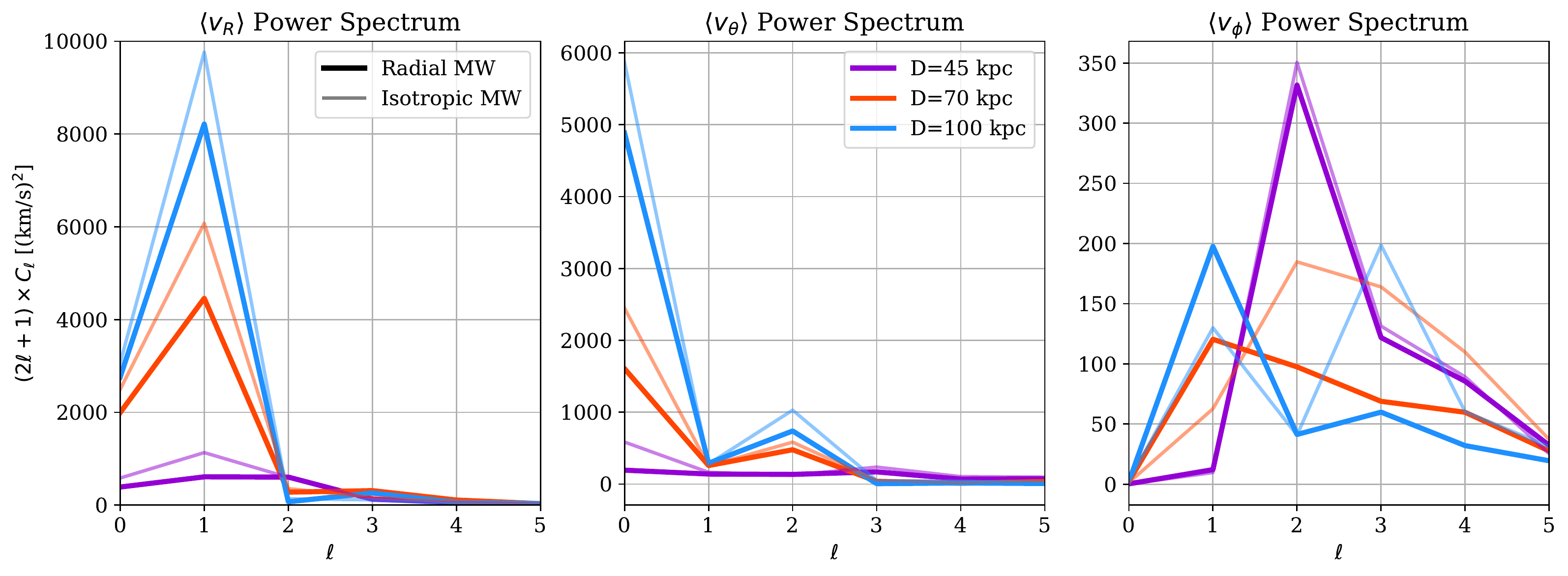}
    \caption{Corresponding power spectra for the velocity maps shown in Figure \ref{fig:maps_3d_v}. Bold lines show the power spectra for the radially anisotropic halo; faded lines show the power spectra for the isotropic halo (maps not shown). The Collective response causes power in $\ell=1$ in $v_{R}$ and $\ell=0$ in $v_{\theta}$, both of which increase as a function of distance. The power spectra illustrate that the Collective response is stronger in the isotropic halo. The Transient response is captured by $\ell=2$ in $v_R$ and $v_{\phi}$.
    We note that the $y-$axis ranges are different for each component of motion, to highlight the differences within each component as a function of distance; in particular, the power in $v_{\phi}$ is much, much lower than the other two components of motion (except at 45 kpc).}
    \label{fig:ps_3dv}
\end{figure*}

The power spectra for these maps are shown by the bold lines in Figure \ref{fig:ps_3dv}. The kinematic patterns are concisely summarized by the angular power spectra. The radial velocity dipole that increases in magnitude with Galactocentric distance is reflected by the increasing power in the $\ell=1$ modes; the increasing mean polar velocity as a function of distance is captured by the increasing power in $\ell=0$ (i.e., the monopole). The power in $v_{\phi}$ is strongest at 45 kpc, where the stars in the Transient response are closest to the present-day position of the LMC and are accelerated by its COM motion.

The faded lines in Figure \ref{fig:ps_3dv} are the resulting power spectra for the GC19 simulation using an isotropic MW halo. As discussed in GC19,
 the Collective response is stronger in the isotropic simulations. 
This is reflected in the resulting power spectra. In the radial velocity power spectrum, at large radii, the magnitude of the velocity dipole is larger for the isotropic halo, as is the magnitude of the $v_{\theta}$ monopole, reflecting the stronger Collective response.

\subsection{LMC Mass Dependence}
\label{sec:lmc_mass}

\begin{figure*}
    \centering
    \includegraphics[width=0.87\textwidth]{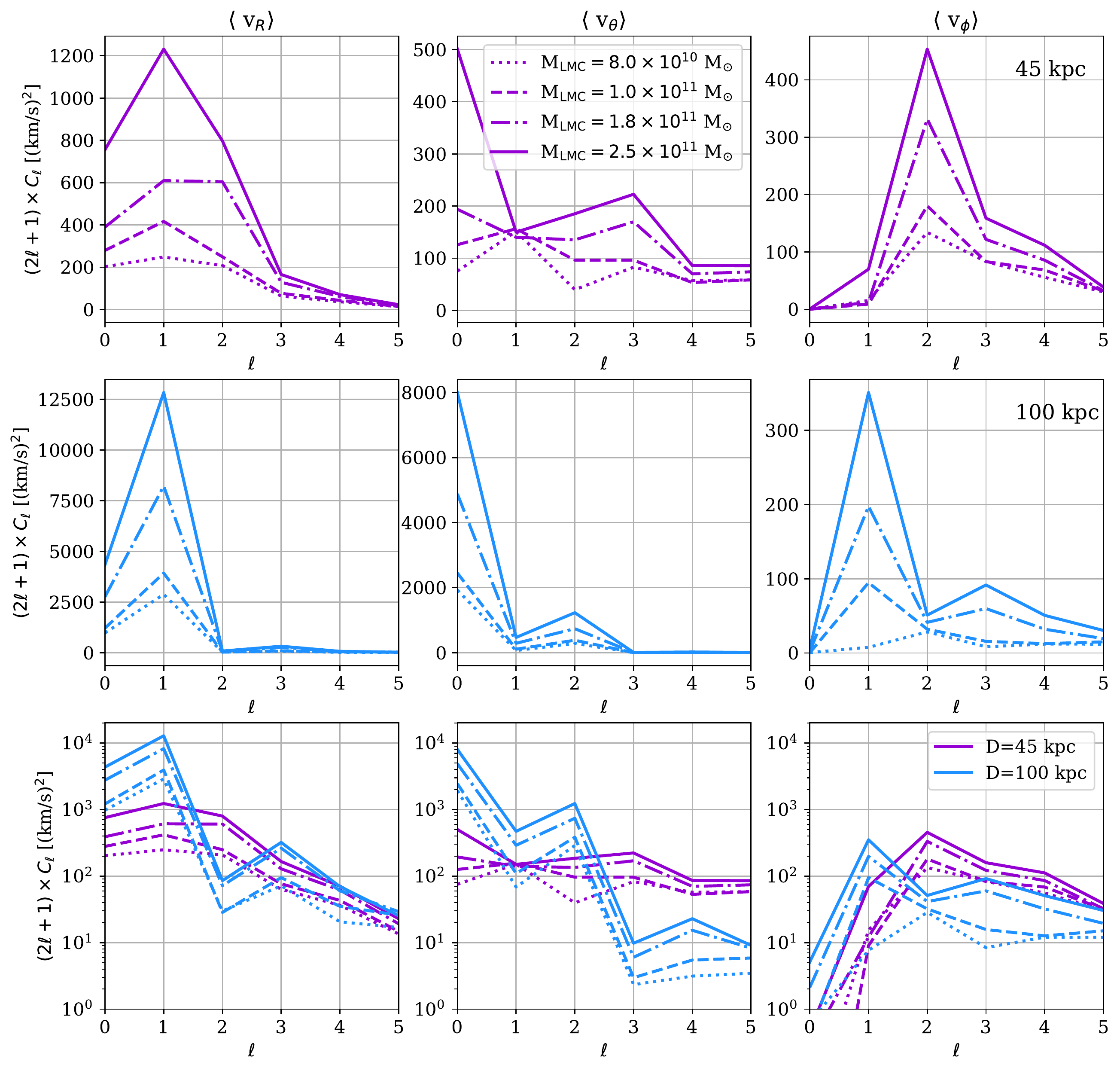}
    \caption{Power spectra for the mean velocities for the GC19 simulations with the radially anisotropic MW model and different LMC masses. Left panels show the angular power spectra for the radial velocity $\langle v_{R} \rangle$; middle panels show the polar velocity $\langle v_{\theta} \rangle$; and right panels show $\langle v_{\phi}\rangle$. The first and second rows show power spectra (plotted linearly) for velocities at computed at 45 kpc and 100 kpc, respectively. Different linestyles show the range of LMC masses simulated in GC19: $M_{\rm LMC}=0.8, 1.0, 1.8, 2.5 \times 10^{11} M_{\odot}$. We emphasize that the $y-$axis ranges are different in each panel, to highlight the differences in the power spectra for the different LMC masses. The spherical harmonic coefficients clearly increase with mass of the LMC.
    Bottom panel: same as first and second rows, but power spectra are plotted on a logarithmic scale. The shape of the power spectra are broadly the same for all the different LMC masses, but scale with LMC mass.}
    \label{fig:ps_mass}
\end{figure*}

Because the mass of the LMC is still very uncertain, GC19 simulated the LMC's infall at four different masses: $M_{\rm LMC,vir}=0.8, 1.0, 1.8, 2.5 \times 10^{11} M_{\odot}$. Figure \ref{fig:ps_mass} shows the resulting power spectra for the mean velocities in the three spherical components of motion at 45 kpc, 70 kpc, and 100 kpc (top, middle and bottom panels, respectively). The different linestyles in each figure represent the different LMC masses.  

While the shape of these power spectra at a given distance for a given component of motion are overall very similar to one another (highlighted by the bottom panel, which shows the power spectra on a logarithmic scale), the total power at a given $\ell$ value clearly trends with LMC mass. In GC19, by quantifying the ``strength" of the wake as the magnitude of the density fluctuations, they find that the strength of the wake is comparable at 45 kpc for all LMC masses (see their Figure 25; though this is for the isotropic MW model, which has a very weak Transient response). Based on the top panels of Figure \ref{fig:ps_mass}, we can see that the power at different $\ell$ values increases strongly with LMC mass even at 45 kpc. We emphasize that the $y-$axis labels are different in each panel, to emphasize the effect of changing the LMC mass; we note that the peak of the power spectrum is largest in $v_R$ at all distances, and the $v_{\phi}$ power spectrum has the least amount of power at all distances. 

The sensitivity of these signals to the LMC mass shown in Figure \ref{fig:ps_mass} emphasizes the significance of the paradigm shift from a  $10^{10} M_{\odot}$ mass LMC to a favored $\sim 10^{11} M_{\odot}$ mass LMC. If the LMC were only $10^{10} M_{\odot}$, this would be a factor of eight less massive than the least massive LMC simulated by GC19; based on the power spectra in Figure \ref{fig:ps_mass}, we can see that the signatures of the DM wake in this scenario would be very weak. Because the LMC is favored to be $\sim 10 \%$ of the MW mass, as opposed to $\sim 1 \%$ (like the other MW satellites), we cannot ignore its gravitational influence on MW halo tracers.

\subsection{Summary}

In summary, low order spherical harmonic expansion is able to capture the salient features in the kinematic patterns that are predicted to arise as a result of the LMC's infall. We have shown that the shape and magnitude of the power spectra depend on the kinematic state of the halo, because the relative strengths of the different wake components depend on the kinematic state of the halo. The overall power is a strong function of the mass of the LMC.

However, one key simplification of the GC19 simulations is that the MW halo model is smooth. The MW stellar halo is known observationally to contain substructure: remnants from disrupted dwarf galaxies, consumed by the MW during its hierarchical formation. In the subsequent section, we investigate how substructure due to accreted dwarf galaxies might obscure the phenomena described in GC19. 

\section{Spherical Harmonic Expansion of Accreted Substructure} 
\label{sec:bj05}

\begin{figure*}
    \centering
    \includegraphics[width=0.8\textwidth]{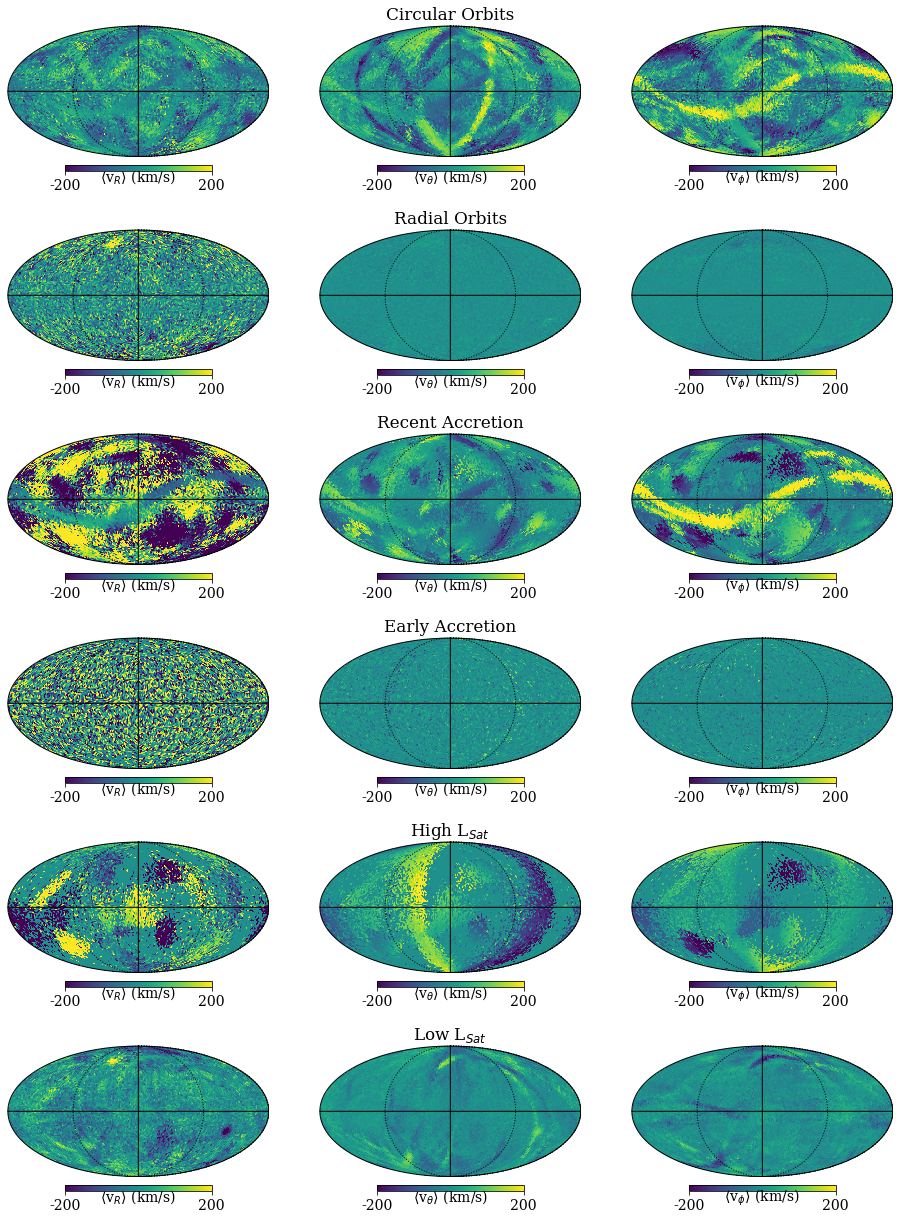}
    \caption{ Velocity maps for the six artificially constructed BJ05 halos, for stars with $30$ kpc$ < r < 50$ kpc. Lefthand panels show radial velocity $v_R$; middle panels show polar velocity $v_{\theta}$; and righthand panels show azimuthal velocity $v_{\phi}$. Pixels in each map are colored by the average velocity. Because of the drastically different accretion histories experienced by these halos, their velocity maps look very different: the halos that experienced only circular, high-$L_{\rm sat}$ and recent accretion events have many more features than the halos that experienced only radial, low-$L_{\rm sat}$ and early accretion events.}
    \label{fig:art_maps}
\end{figure*}

The GC19 simulations model the MW DM halo (and, as result, their stellar halo) as smooth; however, we know that the MW stellar halo is structured. In this section, we investigate how Galactic substructure might complicate our ability to characterize the LMC-induced DM wake using the spherical harmonic expansion of the velocity field, using the \cite{Bullock2005} suite of simulations of purely accreted stellar halos (hereafter BJ05). 

The publicly available BJ05 simulations are \textit{N}-body simulations of accreted dwarf galaxies onto a MW-like parent galaxy. The full suite of simulations consists of 1515 individual accretion events, with a variety of masses, orbital parameters, and accretion times, that together make up the eleven traditional BJ05 halos. Each disrupted satellite is modeled with $10^5$ DM particles. The parent galaxy is represented by a time-evolving potential with disk, halo and bulge components. \cite{Johnston2008} explore in depth how the observable properties of the substructure in these simulations are related to the properties of their satellite progenitors. Here, we explore the links between a galaxy's accretion history, the resulting velocity maps, and the angular power spectra of the velocity field. We note that because there are no DM particles in the parent galaxy halo, there are no density wakes induced in the BJ05 simulations. In this section we are only concerned with spatially varying mean velocities arising from debris from accreted dwarfs. 

Specifically, we consider the six halos with ``artificially constructed" accretion histories discussed in \cite{Johnston2008}. While the standard eleven BJ05 halos have accretion histories from merger trees constructed in $\Lambda$CDM cosmological context, the artificially constructed halos contain debris from accretion events selected from the full BJ05 library that have the desired properties. While all six halos end up with a total luminosity $L \sim 10^9 L_{\odot}$, they assemble their halos very differently. The six artificial halos are:

\begin{itemize}
    \item Circular (Radial) Orbits: halo assembled only from accretion events with $J_{\rm sat}/J_{\rm circ}>0.75$ ($J_{\rm sat}/J_{\rm circ}<0.2$)
    \item Recent (Early) Accretion: halo assembled only from accretion events with $t_{\rm acc}< 8$ Gyr ($t_{\rm acc}>11$ Gyr)
    \item High $L_{\rm sat}$ (Low $L_{\rm sat}$): halo assembled only from accretion events with $L_{\rm sat}>10^7 L_{\odot}$ ($L_{\rm sat}<10^7 L_{\odot}$)
\end{itemize}

\begin{figure*}
    \centering
    \includegraphics[width=\textwidth]{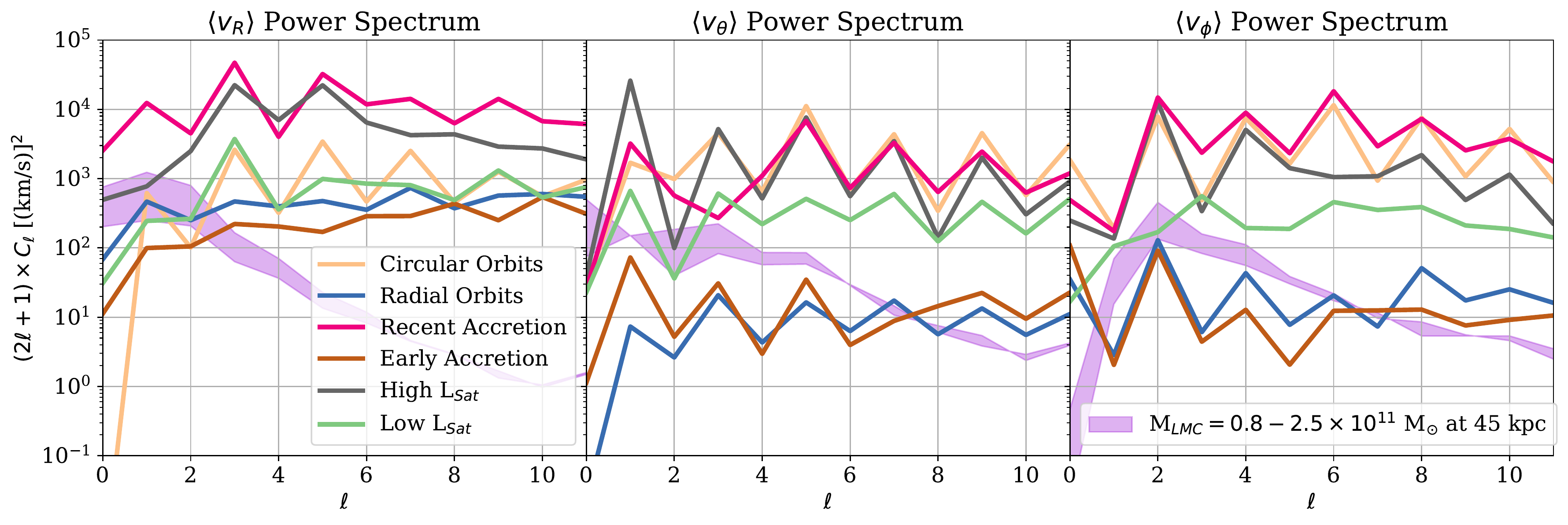}
    \caption{Corresponding angular power spectra for the velocity maps from the BJ05 halos with artificially constructed accretion histories. The purple shaded region indicates the range of power spectra at 45 kpc from GC19, for the full range of simulated LMC masses ($M_{\rm LMC}=0.8-2..5 \times 10^{11} M_{\odot}$). The halos that experienced recent and high-$L_{\rm sat}$ accretion events have more power than the GC19 simulations for nearly all $\ell$. The halo that experienced only circular accretion events also has more power in the tangential components of motion than the GC19 simulations.}
    \label{fig:art_power}
\end{figure*}

Figure \ref{fig:art_maps} shows velocity maps of the six artificially constructed BJ05 halos for stars in the distance range of $30-50$ kpc (with respect to the Galactic center), projected in Mollweide coordinates. As in previous figures, left panels are radial velocity ($v_R$) maps, polar velocity ($v_{\theta}$) maps are in the middle panels, and right hand panels are azimuthal velocity ($v_{\phi}$) maps. We emphasize that the colorbars for these maps range from $[-200, 200]$ \kms, a much larger range than shown for the GC19 simulations; the amplitudes of the velocity fluctuations in these maps are much greater than those due to the LMC-induced DM wake as seen in GC19. As a result, the signatures from substructure are difficult to compare with the wake signatures using the maps alone; the angular power spectra corresponding to these maps, along with the GC19 power spectra, are plotted in Figure \ref{fig:art_power}. We also note that due to the resolution of the BJ05 simulations, there are pixels that contain no star particles; these pixels are assigned to have $\langle v \rangle =0$ \kms, consistent with the assumptions of a equilibrium model.

As a result of the different (and extreme) accretion histories experienced by these halos, their velocity maps look very different. Unsurprisingly, the halo that experienced only early accretion has no features in any components of motion in its velocity maps (bottom panels of Figure \ref{fig:art_maps}), given that all of its accreted material has had sufficient time to become phase mixed. In contrast, the halo that accreted massive, high luminosity satellites has large scale features in all components of motion (second row in Figure \ref{fig:art_maps}). It is also worth noting that this halo accreted $\sim 35\%$ of its mass within the last 8 Gyr (see Figure 7 of \citealt{Johnston2008}); it has therefore experienced recent accretion in addition to massive accretion. The halo that experienced mostly circular accretion events (top row of Figure \ref{fig:art_maps}) has low levels of variation in its radial velocity map, with many thin features of nearly constant (and approximately $0$ \kms) radial velocity, corresponding to streams. These streams have more energy in tangential than radial motion: they appear bands of nearly constant but high velocity in the $v_{\theta}$ and $v_{\phi}$ maps. The halo that experienced only recent accretion (third row of maps in Figure \ref{fig:art_maps}) contains both debris from massive accretion events (as seen by large patches of stars at common velocities) as well as kinematically cold streams (from recent, lower mass, circular events). 

The halos built from radial accretion events and low-luminosity accretion events also do not have large scale features in any components of motion; however, it is important to keep in mind that these halos also have had relatively quiescent recent accretion histories. Both halos had assembled $\sim 95\%$ of their mass 8 Gyr ago (see again Figure 7 of \citealt{Johnston2008}); therefore, any features in these maps are due to relatively recent, low mass events. A radial stream appears as a small bright spot in the $v_R$ maps for both halos; circular streams can be seen as thin bands in the $v_{\theta}$ and $v_{\phi}$ maps for the low-$L_{\rm sat}$ halo.  

The angular power spectra corresponding to these velocity maps are plotted in Figure \ref{fig:art_power}. The halos with the most power at all $\ell$ values are the halo built from recent accretion and the halo built from high luminosity satellites. The halo built from circular accretion events also has high power in $v_{\theta}$ and $v_{\phi}$. The thin streams in the low-$L_{\rm sat}$ halo also result in substantial power ($> 100$ \kms) over many $\ell$ in the three components of motion, though not as much power as the recently accreted, high-$L_{\rm sat}$ and circular halos. 

The halos that experienced only radial accretion events and only early accretion events both have nearly featureless maps in $v_{\theta}$ and $v_{\phi}$; while with few to no features in radial velocity, we note that the $v_R$ maps appear noticeably noisier than the other components of motion. This is a result of the fact that these halos have radially biased velocity anisotropy: their radial velocity dispersions are much greater than their tangential velocity dispersions. The resulting radial velocity maps have greater fluctuations from pixel to pixel than their tangential velocity counterparts. In addition, because there are many fewer particles in these simulations than in GC19 (even though we are looking at a larger radial range in BJ05), the pixel to pixel variation is higher for the BJ05 maps than the GC19 maps. This results in some power in the radial velocity power spectrum, albeit with less power than the other halos, and with hardly any power in the tangential velocity components. 

The purple shaded in region in Figure \ref{fig:art_power} shows the range of power at 45 kpc for the GC19 simulations, from the lowest-mass to the highest-mass LMC. The power from the halo formed through recent accretion and the high-$L_{\rm sat}$ halo is much greater than the power from the LMC-induced DM wake at nearly all $\ell$ in all components of motion. Even the circular and low-$L_{\rm sat}$ halos can have comparable signals to the wake in the tangential components of motion. However, the shape of the power spectrum is very different for Galactic substructure than for the LMC-induced DM wake. The power spectra are characterized by a sawtooth pattern, with peaks at odd $\ell$ values for $v_R$ and $v_{\theta}$ and peaks at even $\ell$ values for $v_{\phi}$. This sawtooth pattern is indicative of the fact that spherical harmonics are not the ideal basis for the velocities of stars in substructure; we explore why the power spectra have these features in the Appendix. 

Based on the power spectra plotted in Figure \ref{fig:art_power}, if the MW stellar halo is dominated by debris from recent, massive accretion events, the signal from the LMC-induced DM wake in the velocity field would be overwhelmed by the Galactic substructure. This signal is from debris that has not yet phase-mixed; based on the velocity maps, we can see that this substructure is clearly visible as overdensities in phase-space, not just in velocity space. Therefore, one could take advantage of the fact that many of these features would be clearly identifiable observationally as overdensities, and could be removed from the analysis relatively easily. 

In addition, while it is likely that debris from an early, massive accretion event dominates the inner halo (i.e., the Gaia Sausage/Gaia-Enceladus, $\sim 10$ Gyr ago; e.g., \citealt{Belokurov2018}, \citealt{Helmi2018}) the current consensus is that the MW has had a fairly quiescent recent accretion history. This consensus has emerged based on numerous studies, including studies of the structure and kinematics of stars of the Galactic disk plane (e.g., \citealt{Gilmore2002}, \citealt{Hammer2007}, \citealt{Ruchti2015}); the steep stellar density profile beyond $\sim 25$ kpc in the halo (e.g., \citealt{Deason2013a}, \citealt{Pillepich2014}, \citealt{Deason2018}); and the amount of substructure in the halo relative to predictions from simulations (e.g., \citealt{Lancaster2019a}). An alternative scenario is one in which the MW has experienced more recent low luminosity or radial accretion events with debris that is harder to find observationally; for example, \cite{Donlon2019} suggest that a recent ($\sim 2$ Gyr), radial merger (with $M \sim 10^9 M_{\odot}$), that mixes efficiently, could explain the Virgo Overdensity \citep{Vivas2001}. Regardless, the MW's accretion history is not believed to be dominated by recent massive accretion. 

The major exceptions to the picture of the MW as having a quiescent (massive) recent merger history are the relatively recent accretion of Sagittarius (Sgr; $\sim 6$ Gyr ago) and the LMC ($\sim 2$ Gyr ago). In the following section, we explore how the presence of debris from Sgr might impact the spherical harmonic expansion of the halo velocity field. 

\section{Sagittarius}
\label{sec:sgr}

\begin{figure*}
    \centering
    \includegraphics[width=0.63\textwidth]{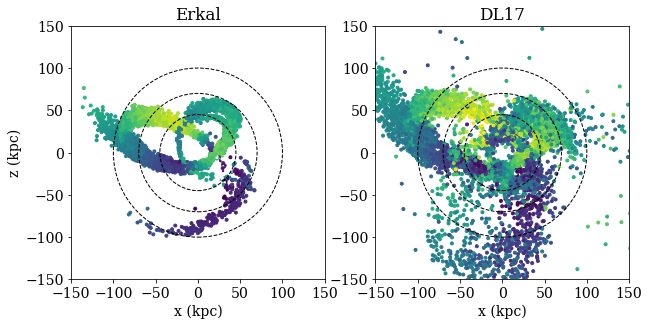}
    \includegraphics[width=0.1\textwidth]{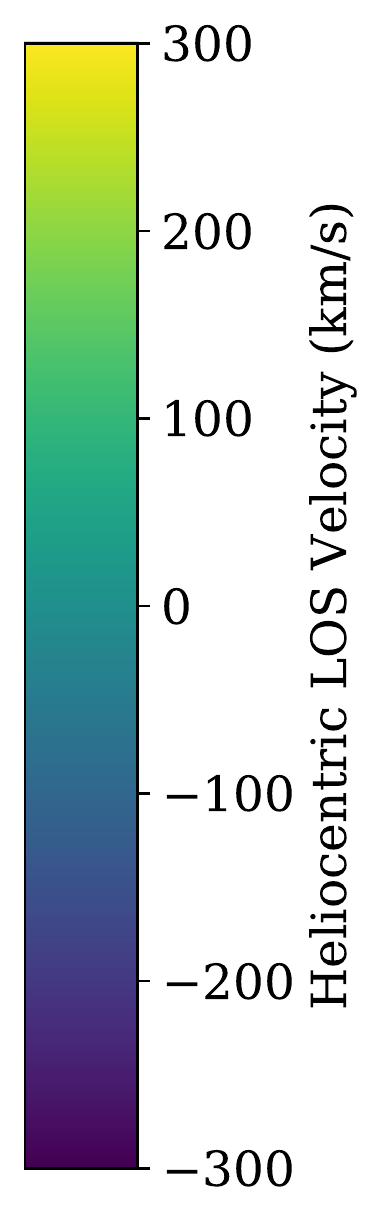}
    \includegraphics[width=0.33\textwidth]{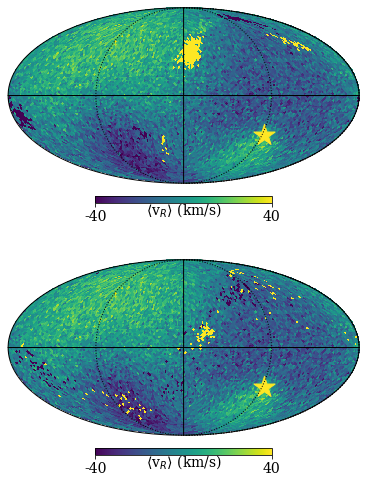}
    \includegraphics[width=0.65\textwidth]{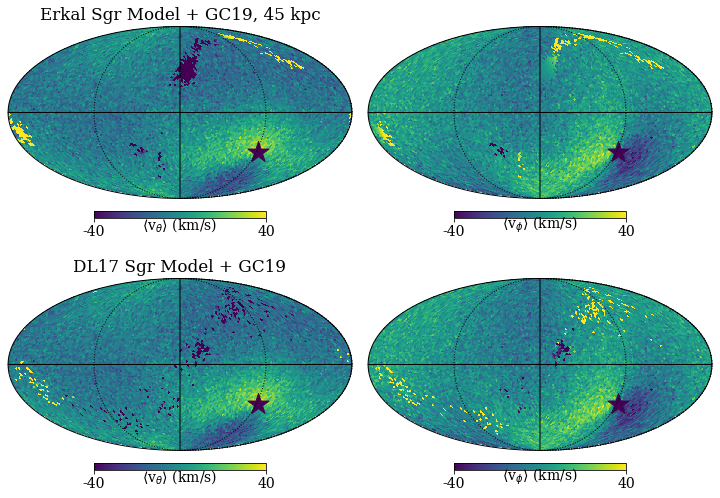}
    
    \caption{Top panels: positions in the $x-z$ plane, color coded by heliocentric LOS velocity, for stars in the Erkal Sgr model (left) and the DL17 Sgr Model (right). Dashed lines indicate 45 kpc, 70 kpc, and 100 kpc. Middle panels: velocity maps for the Erkal Sgr model overlaid onto the GC19 stellar halo, in a 5 kpc shell centered at 45 kpc. The angular position of the LMC is indicated by the star symbol, color coded to show the sign of the LMC's velocity in each component of motion. We note that we have restricted the velocity colorbar to $\pm 40$ \kms, so that the wake signatures are still visible by eye; as can be seen in the colorbar in the top panels, the range of velocities of stars in Sgr is much greater than the range of mean velocities in the GC19 halo at 45 kpc. In addition, we note that the overdense region in the North in these maps is not the Sgr progenitor, but rather part of the leading arm. Lower panels: same as middle panels, but for the DL17 Sgr model overlaid on the GC19 halo.}
    \label{fig:sgr_pos}
\end{figure*}

In Section \ref{sec:lmc_only}, we discussed in detail the spherical harmonic expansion of the kinematic variation that arises due to the infall of the LMC. In Section \ref{sec:bj05}, we saw that recent, massive accretion events can cause kinematic variation on large scales, which in turn can result in substantial power over many $\ell$ values in the power spectrum. In the MW, the debris from the most recent, massive accretion event (aside from the LMC) is found in the Sgr stream (\citealt{Ibata2001}). The progenitor of Sgr was a relatively massive, luminous satellite ($L\sim 10^8$, $M > 10^9 M_{\odot}$; e.g., \citealt{Penarrubia2010}, \citealt{Niederste-Ostholt2012}, \citealt{Deason2019}; \citealt{Gibbons2017} suggest an even higher mass $M>6 \times 10^{10}$). Debris from the Sgr accretion event is found all over the sky over Galactocentric distances ranging from $\sim 15$ kpc to $\sim 130$ kpc (e.g., \citealt{Majewski2003}, \citealt{Belokurov2014},  \citealt{Hernitschek2017}, \citealt{Sesar2017}). In this section, we investigate how the presence of Sgr stars might obscure the signal from the LMC-induced DM wake in the velocity field. 

We consider two different models for the Sgr stream. The first Sgr model we consider is a fit of the Sagittarius stream in the presence of the LMC which we dub the Erkal model. This model uses the same stream fitting machinery of \cite{Erkal2019b} which accounts for the reflex motion of the Milky Way due to the LMC. This technique rapidly generates streams using the modified Lagrange Cloud stripping technique from \cite{Gibbons2014}. For this model, we fit the radial velocity and distances from \cite{Belokurov2014} and on-sky positions from \cite{Belokurov2006,Koposov2012} for the bright stream. 

Motivated by the results of \cite{Law2010}, we model Sgr as a $2.5\times10^{8} M_\odot$ Plummer sphere with a scale radius of $0.85$ kpc. The progenitor is rewound for 5 Gyr in the combined presence of the Milky Way and LMC and then disrupted to the present to form the Sgr stream. For the Milky Way potential, we take the triaxial NFW generalization from \cite{Bowden2013} which allow for different inner and outer density flattenings. We fix the concentration to $c=15$. As a further generalization, we allow for an arbitrary rotation of this triaxial halo so that its axes are not necessarily aligned with the galactic Cartesian coordinates. We also include a similar disk and bulge to the \texttt{MWPotential2014} from \cite{Bovy2015}: a Miyamoto-Nagai disk \citep{Miyamoto1975} with a mass of $6.8\times10^{10} M_\odot$ with a scale radius of $3$ kpc and a scale height of $0.28$ kpc, and a Hernquist bulge \citep{Hernquist1990} with a mass of $5\times10^9 M_\odot$ and a scale radius of 0.5 kpc. We use the dynamical friction prescription of \cite{Jethwa2016} both for the dynamical friction from the Milky Way on the LMC and from the Milky Way on Sgr. The distance, radial velocity, and proper motions of the Sgr are left as free parameters with priors set by observations \citep{McConnachie2012}. For the LMC, we give it a fixed position and velocity based on its mean observed distance \citep{Pietrzyski2013}, radial velocity \citep{vanderMarel2002}, and proper motion \citep{Kallivayalil2013}. We model the LMC as a Hernquist profile \citep{Hernquist1990} with a scale radius of 25 kpc and a free mass with a uniform prior from $0-3\times10^{11} M_\odot$. In order to account for the fact that Sgr was initially more massive and had a substantial dark matter component which would have experienced more dynamical friction, we have an additional free parameter that increases the mass of Sgr by $\lambda_{\rm DF}$ when computing its dynamical friction. This has a uniform prior between 0 and 20. Thus, all together we have 15 free parameters: the mass and scale radius of the NFW profile ($M_{\rm NFW}, r_{s\,{\rm NFW}}$), an inner and outer minor and intermediate axis flattening ($q_0,p_0,q_\infty,p_\infty$), three angles to describe the rotation of the triaxial halo, the mass of the LMC ($M_{\rm LMC}$), the mass multiplier $\lambda_{\rm DF}$, and finally the proper motions, radial velocity, and distance of Sgr progenitor at the present day.

We use the MCMC package from \cite{Foreman-Mackey2013} to estimate the parameters of Sgr. We use 100 walkers for 2000 steps with a 1000 step burn in. The best-fit parameters require an LMC mass of $2.0\times10^{11} M_\odot$, flattenings of $q_0 = 0.68, p_0 = 0.87, q_\infty = 0.81, p_\infty = 0.94$. The mass multiplier $\lambda_{\rm DF}=9.5$ suggesting that fitting the Sgr stream requires more dynamical friction than the low mass we have assumed would provide. The best-fit Milky Way mass is $6.76\times10^{11} M_\odot$ with a scale radius of $15.3$ kpc. Although this Milky Way mass is relatively modest, we note that the scale radius is also quite small. Despite the flexibility of this model, we note that it does not perfectly match the distance but most importantly for this work, it gives a good match to the radial velocity across the sky. A comparison of the model with radial velocities from \cite{Belokurov2014} is shown in Figure \ref{fig:sgr_vgsr}. The positions of the stars for this Sgr model in the  $x-z$ plane, color coded by heliocentric line-of-sight velocity, are shown in the top left panel of Figure \ref{fig:sgr_pos}.

The second Sgr model we consider is the publicly available model from \cite{Dierickx2017} (hereafter DL17).\footnote{https://mdierick.github.io/project2.html} The $x-z$ positions of stars from this model, again color-coded by heliocentric line-of-sight velocity, are shown in the righthand panel of Figure \ref{fig:sgr_pos}. In DL17, they first utilize a semi-analytic approach to derive initial conditions for the Sgr progenitor, by integrating the equations of motion forward in time over 7-8 Gyr and comparing the resulting position and velocity vector to the observed properties of the Sgr remnant. They assume virial masses $M_{\rm Sgr}=10^{10} M_{\odot}$ and $M_{MW}= 10^{12} M_{\odot}$. To then model the disruption of Sgr, they run an \textit{N}-body simulation using the derived initial conditions from the semi-analytic approach, modelling both a live MW and Sgr. While this model does reproduce many of the features of the Sgr stream, including the positions of stars observed in 2MASS (\citealt{Majewski2004}) and SDSS (\citealt{Belokurov2014}), and the large apocentric distances observed in \cite{Sesar2017}, we emphasize that the \textit{N}-body simulation is not tuned to fit the observations of the Sgr stream. As a result, certain properties of the stream (for example, the LOS velocities along the leading arm) are not well matched by the data. 

To investigate how stars from Sgr might impact the power spectrum of the MW halo's velocity field, we overlay the two models for the Sgr stream on to the fiducial anisotropic GC19 simulation (with $M_{LMC}=1.8 \times 10^{11} M_{\odot}$). To combine the two independent simulations, we assign the total Sgr stellar mass to be 10$\%$ of the total stellar mass of the GC19 halo. This ratio is consistent with current estimates of the total stellar mass of the MW ($\sim 10^9 M_{\odot}$; e.g., \citealt{Deason2019}, \citealt{Mackereth2020}) and Sgr ($M_{\rm Sgr, *} \sim 10^8 M_{\odot}$; e.g., \citealt{Deason2019}, \citealt{Niederste-Ostholt2012}). 

\begin{figure*}
    \centering
    \includegraphics[width=\textwidth]{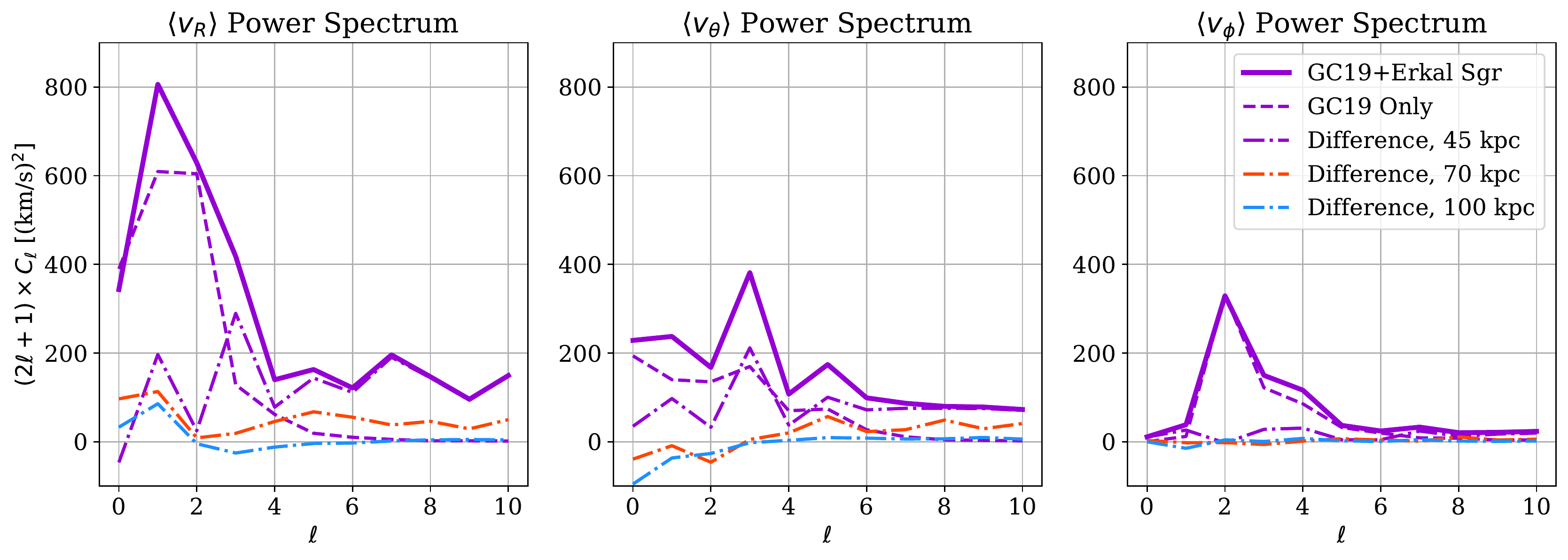}
    \includegraphics[width=\textwidth]{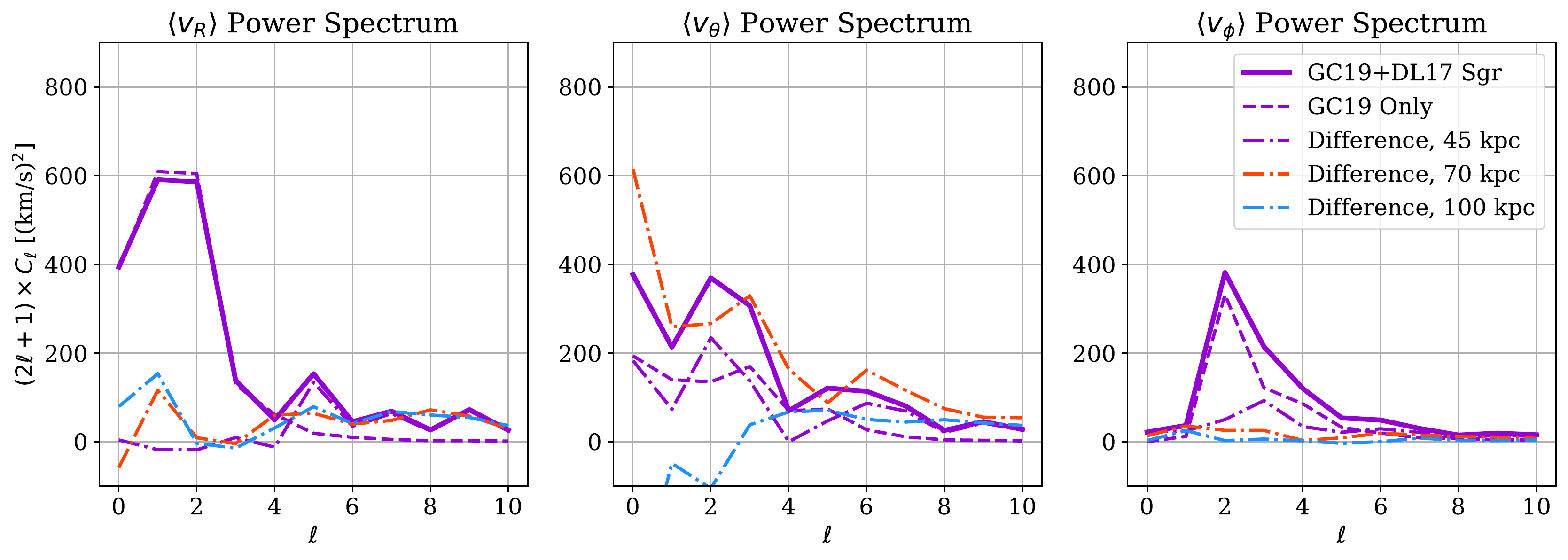}
    \caption{Power spectra for the Erkal Sgr model (top panels) and the DL17 Sgr model (lower panels) overlaid onto the GC19 anisotropic simulation with $M_{LMC}=1.8 \times 10^{11} M_{\odot}$. Solid lines show the resulting power spectra at 45 kpc when the two simulations are combined; dashed lines show the power spectra from GC19 simulation alone at 45 kpc. The difference between the resulting power spectra are plotted as dot-dashed lines, at 45 kpc (purple), 70 kpc (orange), and 100 kpc (blue). The power at low $\ell$ (i.e., large spatial scales), remains generally dominated by the signatures of the LMC-induced DM wake, especially at larger distances (though at 45 kpc, the power due to Sgr in $v_{\theta}$ is comparable to the power due to the wake). However, Sgr does substantially contribute to modes that are sensitive to the LMC-induced wake (e.g., $\ell=1$ in $v_R$, $\ell=0$ in $v_{\theta}$); the influence of Sgr stars should therefore be modeled if quantifying the strength of the wake using SHE.}
    \label{fig:sgr_ps}
\end{figure*}

The resulting velocity maps at 45 kpc of the two Sgr models overlaid on the GC19 simulations are shown in the middle panels (for the Erkal model) and lower panels (for the DL17 model) of Figure \ref{fig:sgr_pos}. The corresponding power spectra for the velocity maps in Figure \ref{fig:sgr_pos} are shown by the thick solid lines in Figure \ref{fig:sgr_ps}. We only show the full power spectra at 45 kpc, as Sgr does not substantially contribute to the overall power spectrum at larger radii (with the exception of the DL17 model at 70 kpc; this results from stars accelerating towards and away from the stream apocenters, which are at larger distances for the DL17 model than for the Erkal model). From left to right, power spectra for $v_R, v_{\theta}, v_{\phi}$ are plotted; the thick solid lines are the power spectra from the combination of the GC19 simulation with the Sgr models (top panels show the results when using the Erkal model; lower panels show the results from the DL17 model) at 45 kpc. The dashed line shows the power spectrum from the GC19 simulation (excluding Sgr). Dotted dashed lines show the difference between the halo including Sgr and excluding Sgr (i.e., the contribution of Sgr to the overall power spectrum), at 45 kpc (purple), 70 kpc (orange), and 100 kpc (blue), computed in 5 kpc shells. 

The dot-dashed lines in Figure \ref{fig:sgr_ps}, representing the contribution to the power spectrum due to Sgr, have a similar morphology to the power spectra discussed in Section \ref{sec:bj05} and in the Appendix: they are characterized by a saw-tooth pattern, with peaks at odd $\ell$ values in $v_R$ and $v_{\theta}$. Figure \ref{fig:sgr_ps} shows that the two Sgr models result in different signatures. For the Erkal model, including Sgr increases the peak at $\ell=1$ in $v_R$, while the $v_{\phi}$ power spectrum is mostly unaffected. The DL17 model hardly affects the low $\ell$ power in $v_R$, while the power in $v_{\phi}$ is slightly enhanced. Both models result in higher power in $v_{\theta}$ at all $\ell$; at 45 kpc, $v_{\theta}$ is the only component of motion for which the signal from Sgr is comparable to the signal of the LMC-induced DM wake.

While including Sgr does increase the overall power in orders $\ell$ that contain signatures of the LMC-induced DM wake, the features sensitive to the Collective response (the $\ell=1$ peak in $v_R$ as well as the monopole $\ell=0$ peak in $v_{\theta}$) are stronger at all distances than the power due to both Sgr models alone (the dot-dashed lines in Figure \ref{fig:sgr_ps}; though we note that the DL17 model does substantially increase the power of the monopole in $v_{\theta}$). In addition, the signatures from the Collective response increase as a function of distance; the overall power from Sgr generally decreases as a function of distance (with the exception of the increase in signal at 70 kpc in the DL17 model in $v_{\theta}$). While the signal from Sgr in these key $\ell$ values does not overwhelm the signal from the LMC, it does contribute to the overall power in the orders sensitive to the wake signatures at 45 kpc. Sgr also contributes power in specific $l,m$ modes that are sensitive to the LMC-induced DM wake (e.g., $l=1, m=0$ in $v_R$); however, the phases of the different coefficients are not strongly affected by Sgr in the low $\ell$ values. Modeling the influence of the inclusion of Sgr will be essential in quantifying the strengths of the LMC-induced wake components observationally at smaller Galactocentric distances.

In summary, while Sgr stars will affect the power spectra of the MW halo velocity field at smaller Galactocentric distances, the signatures from the LMC-induced DM wake as predicted by GC19 should still be distinguishable from the Sgr signatures. The primary signature of the Transient response (power in $\ell=2$ in $v_{\phi}$) is largely unaffected by the inclusion of Sgr for both models. In $v_R$ and $v_{\phi}$, the dominant features in the power spectra that arise due to the LMC-induced DM wake are stronger than the features arising due to the inclusion of Sgr, and the power spectra have different morphologies. Because Sgr stars are likely to contribute power at $\ell=1$ in $v_{R}$ and $\ell=0$ in $v_{\theta}$, 
modeling Sgr will be important in characterizing the Collective response at closer distances in the halo ($\sim <50$ kpc). 

In addition, Sgr has been extensively studied, and our knowledge of its velocity structure is better known than ever before with the release of Gaia DR2 (e.g., \citealt{Antoja2020}, \citealt{Ramos2020}, \citealt{Ibata2020}). Like the substructure studied in BJ05, the debris from Sgr is also well known to be overdense on the sky, and many halo studies remove stars believed to be associated with the stream. Therefore, given that we have shown that the wake signatures should still be identifiable even if all Sgr stars are included in the analysis, the prospects for observing these signatures only improve if debris from Sgr is subtracted, even if this subtraction is imperfect.

However, we note that we have only considered the effects of including Sgr stars in the analysis of the stellar halo, and not the effects the infall of the Sgr dwarf may have had on the MW disk and halo over the last $\sim 6-8$ Gyr. Depending on the assumed mass of Sgr, it may have resulted in a substantial shift in the MW barycenter and also caused a wake in the MW DM halo (\citealt{Laporte2018b}). GC19 compare the magnitude of the density perturbations that arise due to Sgr and the LMC using the \cite{Laporte2018b} simulations, and find that the contribution from Sgr is negligible (see GC19's Figure 26); however, they did not discuss the perturbations to the velocity field. We leave the exploration of these effects to future work.  

\section{Conclusions}
\label{sec:concl}

In this paper, we use spherical harmonic expansion to describe the perturbed velocity fields of the MW as a result of the LMC's infall using the simulations from GC19. We explore the ways in which Galactic substructure might obscure the signatures from the wake in the power spectrum, using the BJ05 simulations as well as two models for the Sgr stream. We summarize our primary findings as follows:

\begin{enumerate}
\item{We study the perturbation to the velocity field caused by the LMC-induced DM wake using the simulations from GC19. We find that low-order spherical harmonic expansion of the velocity field in these simulations usefully captures the salient features of the LMC-induced DM wake. We found that increasing power with Galactocentric radius in $\ell=1$ in $v_R$ and $\ell=0$ in $v_{\theta}$ are signatures of the Collective response. At 45 kpc (near the LMC), power in $\ell=2$ in $v_R$ and $v_{\phi}$ are signatures of the Transient response. We find that the amplitude of the power spectra scale with LMC mass.}

\item{We investigate how Galactic substructure might affect the angular power spectrum of the MW's velocity field, using the BJ05 simulations with artificially constructed accretion histories. We find that massive, recent accretion causes large scale, high amplitude fluctuations in the velocity field. Velocity substructure arising due to debris from recent, massive satellites creates much more power in the power spectrum of the velocity field than the perturbation due to the LMC. However, the MW is not believed to have experienced much recent, massive accretion, with the exceptions of Sgr and the LMC itself.}

\item{Given that Sgr is the most recent, massive accretion event experienced by the MW (with the exception of the LMC), we investigate how Sgr stars could impact measurements of the overall MW power spectra and our ability to measure the signatures associated with the LMC-induced DM wake. The power spectrum on large scales (i.e., low $\ell$) remains generally dominated by the signatures from the LMC-induced DM wake; this result complements the GC19 findings that the amplitude of the density wake induced by the LMC's infall is much greater than the density wake induced by Sgr. In addition, overall power due to Sgr decreases as a function of distance, in contrast to the Collective response signatures. However, including Sgr stars does increase the power in modes that are sensitive to the Collective response, especially at 45 kpc. Care should therefore be taken to model the impact of Sgr stars on the power spectrum in studies attempting to use this method for detecting and quantifying the Collective response.}
\end{enumerate}

Based on our findings, performing spherical harmonic expansion in the MW velocity field could be a method for identifying and characterizing the LMC-induced DM wake, which would in turn provide constraints on the mass and orbital history of the LMC. There remain technical challenges associated with implementing this technique with observational data (e.g., incorporating measurement uncertainties, limited sky coverage of spectroscopic programs, combining data from different surveys); we leave a detailed exploration of how to estimate the spherical harmonic expansion coefficients from realistic data to future work. However, the future is bright given the upcoming observational programs that will map our Galaxy's phase-space structure. The first two Gaia data releases have already transformed our understanding of the MW's kinematic structure; with future releases from Gaia mission in conjunction with the Gaia spectroscopic follow-up programs (e.g., DESI, 4MOST, WEAVE), as well as future astrometric (e.g., Rubin Observatory Legacy Survey of Space and Time, WFIRST) and spectroscopic programs (e.g., SDSS-V MW Mapper), the halo velocity field will be better known than ever before. 

While the GC19 only include the MW and the LMC, they reveal the complex behavior that arises in the phase space structure of the MW halo due to the infall of the LMC. However, many complications remain to be addressed. While we explored how Galactic substructure might obscure the signals from the wake, the assumption of smoothness of the MW halo in GC19, as well as their mapping of the stellar halo on to the DM halo, remain important to keep in mind. While much of the stellar halo is phase-mixed in the inner regions of the MW, at larger distances (e.g., $r>50$ kpc), we may need to rely on debris that is not phase mixed in order to see wake signatures. In addition, the stellar halo may not be in equilibrium with the DM halo for several reasons. First of all, simulations show that accretion is not the only mechanism by which stellar halos form: simulated halos show substantial fractions of stars that formed in-situ (e.g., \citealt{Zolotov2009}, \citeyear{Zolotov2010}). \cite{Yu2020} show that in-situ stars (formed in outflows of the host galaxy) can be ejected to large distances in the halo, and can comprise a substantial fraction (5-40\%) of the stars in outer halos (50-300 kpc). \cite{Bonaca2017} use the kinematics of stars from the \textit{Gaia} data release to argue that the MW halo has an in-situ component. Second, because stars are more concentrated within halos than DM, the DM is preferentially stripped initially (e.g., \citealt{Smith2016}). Third, the dynamical mass-to-light ratios of dwarf galaxies have been observed to vary over orders of magnitude (e.g., \citealt{McConnachie2012}): the relative amounts of mass accreted in stars and DM is therefore not constant. 
Finally, a significant fraction of the DM accretion is ``smooth'' (e.g. \citealt{Angulo2010}; \citealt{Genel2010}), in contrast to the relatively ``lumpy'' accretion that builds up the stellar halo. While methods have been developed using cosmological simulations to determine DM halo distributions based on stellar halo distributions (\citealt{Necib2019}), these complex effects were not included in the creation of stellar halos from DM halos in GC19. In addition, while GC19 varies the mass of the LMC, they assume a single mass for the MW and do not simulate a range of orbital histories. 

Many of these complications can be addressed by studying DM wakes more generally in cosmological simulations. These simulations contain both in-situ and accreted halo components and have experienced a variety of accretion histories. By applying this technique to encounters between MW-like galaxies and massive satellites in cosmological simulations, we can identify wake signatures for a range of mass ratios and orbital histories. In the present era of wide-field 3D kinematic datasets coupled with detailed high resolution simulations, we can move beyond equilibrium models and develop new methods for characterizing the complex processes that have shaped our Galaxy's formation. 

\acknowledgements{ We thank the anonymous referee for their helpful comments and suggestions.
ECC and RL are supported by Flatiron Research Fellowships at the Flatiron Institute. The Flatiron Institute is supported by the Simons Foundation. AD is supported by a Royal Society University Research Fellowship. KVJ's contributions were supported by NSF grant AST-1715582. NGC and GB acknowledge support from HST grant AR 15004 and NASA ATP grant 17- ATP17-0006. CFPL's contributions are supported by world premier international research center initiative (WPI), MEXT, Japan. This project was developed in part at the 2019 Santa Barbara Gaia Sprint, hosted by the Kavli Institute for Theoretical Physics at the University of California, Santa Barbara. This research was supported in part at KITP by the Heising-Simons Foundation and the National Science Foundation under Grant No. NSF PHY-1748958. ECC would like to thank Andrew Wetzel, David Hogg, Adrian Price-Whelan and David Spergel for helpful scientific discussions. 

\software{Astropy (\citealt{astropy:2013}, \citeyear{astropy:2018}), Healpy \citep{Zonca2019}, IPython \citep{Perez2007}, Matplotlib \citep{Hunter2007}, Numpy (\citealt{oliphant2006guide}, \citealt{vanderWalt2011}), Pandas \citep{mckinney-proc-scipy-2010}, Scipy \citep{2020SciPy-NMeth}, Starry \citep{Luger2019}}}

\begin{appendix}

\begin{figure*}[b]
    \centering
    \includegraphics[width=\textwidth]{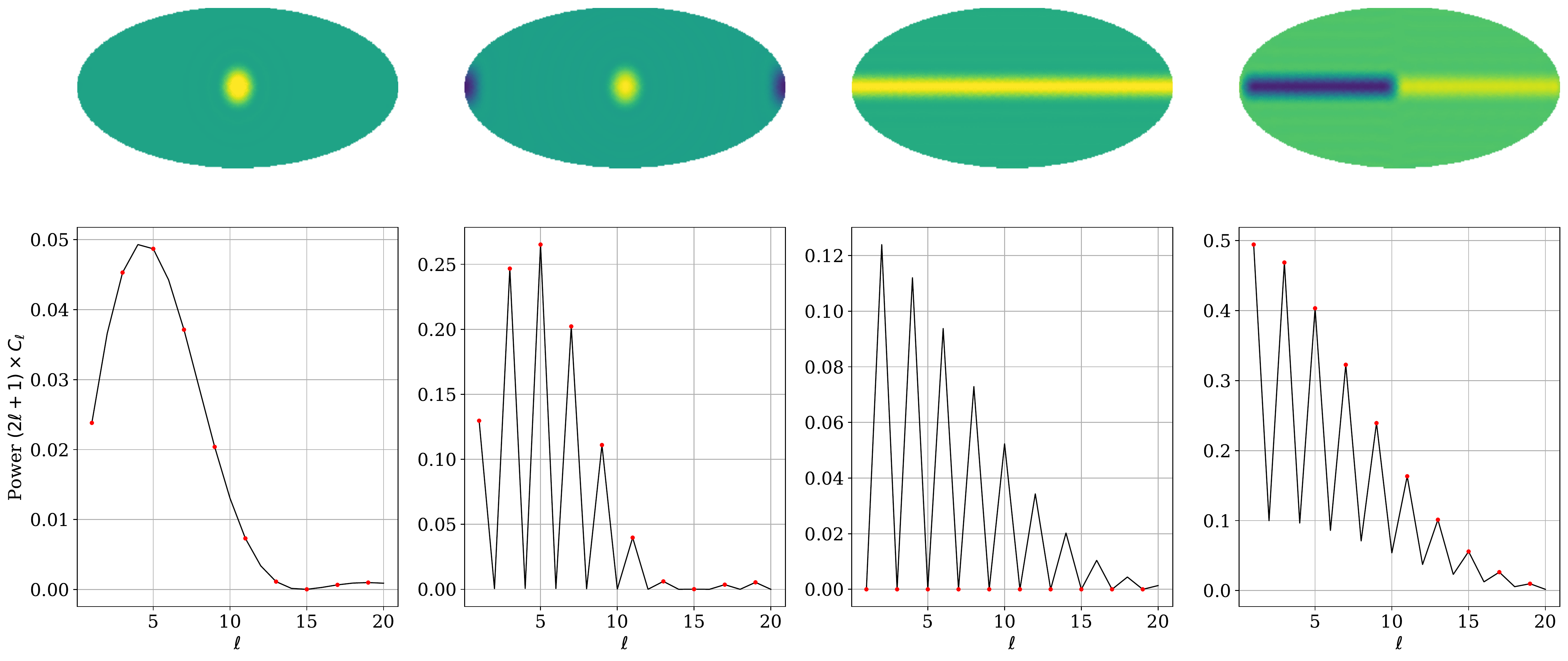}
    \caption{Top panels show intensity maps generated by \texttt{starry} for four different patterns, plotted in Mollweide projection.  Bottom panels show the corresponding power spectra. In the power spectra, the power at odd $\ell$ values is shown in red. While a single spot generates a rather smooth spectrum as a function of $\ell$, when there are two spots of opposite signs across from one another in the map, the resulting power spectrum shows spikes at odd $\ell$. Banding can also cause spikes, but at even $\ell$ (third panel); if the band is weighted by a dipole, then the spikes occur at odd $\ell$ (rightmost panels).}
    \label{fig:starry}
\end{figure*}

In Section \ref{sec:bj05}, we see that the power spectra for all of the BJ05 halos with debris that is not yet phase-mixed are characterized by a sawtooth pattern, with peaks at odd $\ell$ values for $v_R$ and $v_{\theta}$ and peaks at even $\ell$ values for $v_{\phi}$. To understand why these features in the power spectra are arising, we show the resulting power spectra from several simple intensity maps using the software \texttt{starry} (\citealt{Luger2019}) in Figure \ref{fig:starry}. The top panels of Figure \ref{fig:starry} show intensity maps, while the bottom panels show the corresponding power spectra. While a single spot gives rise to a rather smooth power spectrum, with power in even as well as odd $\ell$ values, two spots of opposing sign (located 180$\degree$ apart from one another) result in a power spectrum with peaks at odd $\ell$ values. As seen in the right panels, banding can also give rise to the sawtooth pattern: banding at constant intensity yields peaks at even $\ell$ values, whereas a band weighted by a dipole creates peaks at odd $\ell$ values. 

\begin{figure*}
    \centering
    \includegraphics[width=\textwidth]{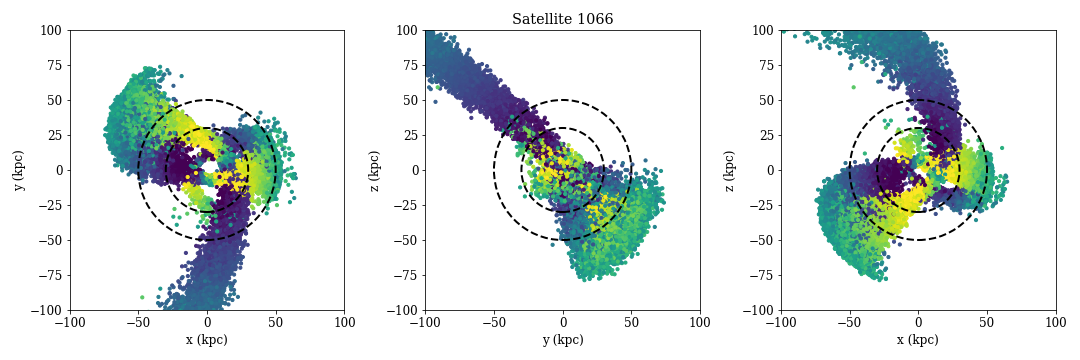}
    \includegraphics[width=\textwidth]{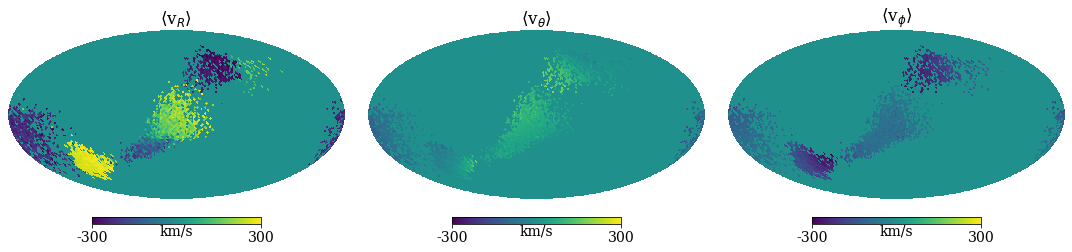}
    \includegraphics[width=\textwidth]{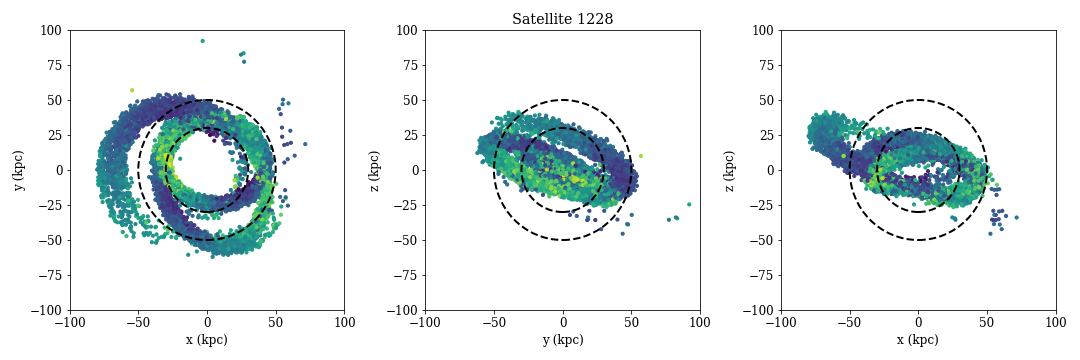}
    \includegraphics[width=\textwidth]{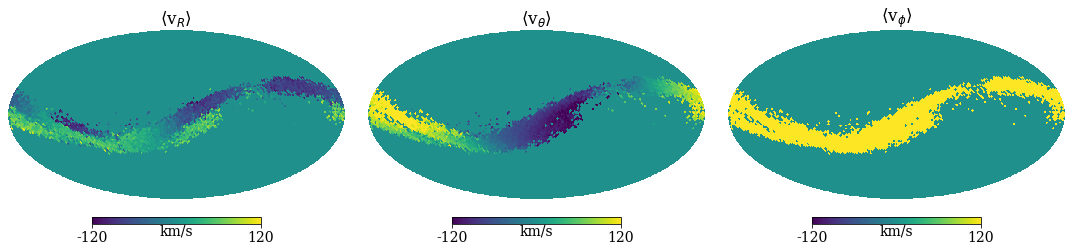}
    \caption{Top panels: $(x,y,z)$ positions for all stars originating from BJ05 satellite 1066 ($M_{\rm Sat}=9.1 \times 10^{10}~M_{\odot},t_{\rm Sat}=4.58~\rm{Gyr}, J_{\rm Sat}/J_{\rm circ}= 0.68$). Points are color coded by their radial velocities, over the range $[-300, 300]$ \kms. Dashed lines indicate $30-50$ kpc. Second row: velocity maps, for stars in the radial range of $30-50$ kpc, in the three components of motion in spherical coordinates. This massive, recent accretion events creates shells in the halo with strong velocity gradients. When we take a cross section of this substructure, we see ``spots" in the $v_R$ maps of opposite signs that are approximately separated by $180 \degree$. Lower panels: same as top panels, but for BJ05 satellite 1228 ($M_{\rm Sat}=5.54 \times 10^{10}~M_{\odot},t_{\rm Sat}=6.55~\rm{Gyr}, J_{\rm Sat}/J_{\rm circ}= 0.99$). Points are color coded by their radial velocities, over the range $[-120, 120]$ \kms. This circular, recent accretion event creates a stream, which create banding features in the velocity maps.}
    \label{fig:bj_sat1066}
\end{figure*}

To understand why the halos have these features in their velocity maps and power spectra, we explore the properties of two specific accretion events. First we consider BJ05 satellite 1066 ($M_{\rm Sat}=9.1 \times 10^{10}~M_{\odot},t_{\rm Sat}=4.58~\rm{Gyr}, J_{\rm Sat}/J_{\rm circ}= 0.68$). This halo is accreted both by the recently accreted BJ05 halo and the high-$L_{\rm sat}$ halo. The top panels of Figure \ref{fig:bj_sat1066} show the positions of all stars from this satellite in Cartesian coordinates, color coded by their radial velocities. Dashed lines indicate the radial range of $30-50$ kpc; velocity maps in spherical coordinates are shown in the second row of panels. This massive, recently accreted satellite has formed large shell-type features. When we only consider stars from this satellite that are located between 30-50 kpc from the center, we see that our cross section does not include the apocenters of these shells (where stars have $v_R \sim 0 $ \kms), but rather stars that are moving quickly either towards an apocenter (yellower spots) or away from apocenter back towards the center of the galaxy (bluer points). In the radial velocity map, the patches of stars with opposite velocity are separated by approximately 180 $\degree$ from each other in this projection. This accretion event has a much narrower range of velocities in $v_{\theta}$ and $v_{\phi}$; the $v_{\theta}$ map has a fairly smooth continuum, going from positive $v_{\theta}$ as the stars moves towards and away the first apocenter to negative $v_{\theta}$ as the stars pass through the second apocenter. This results in banding weighted by a dipole, a version of the map shown to the rightmost panel in Figure \ref{fig:starry}. In $v_{\phi}$, the velocities are always negative; the velocity map appears as a relatively constant band. 

\begin{figure*}
    \centering
    \includegraphics[width=\textwidth]{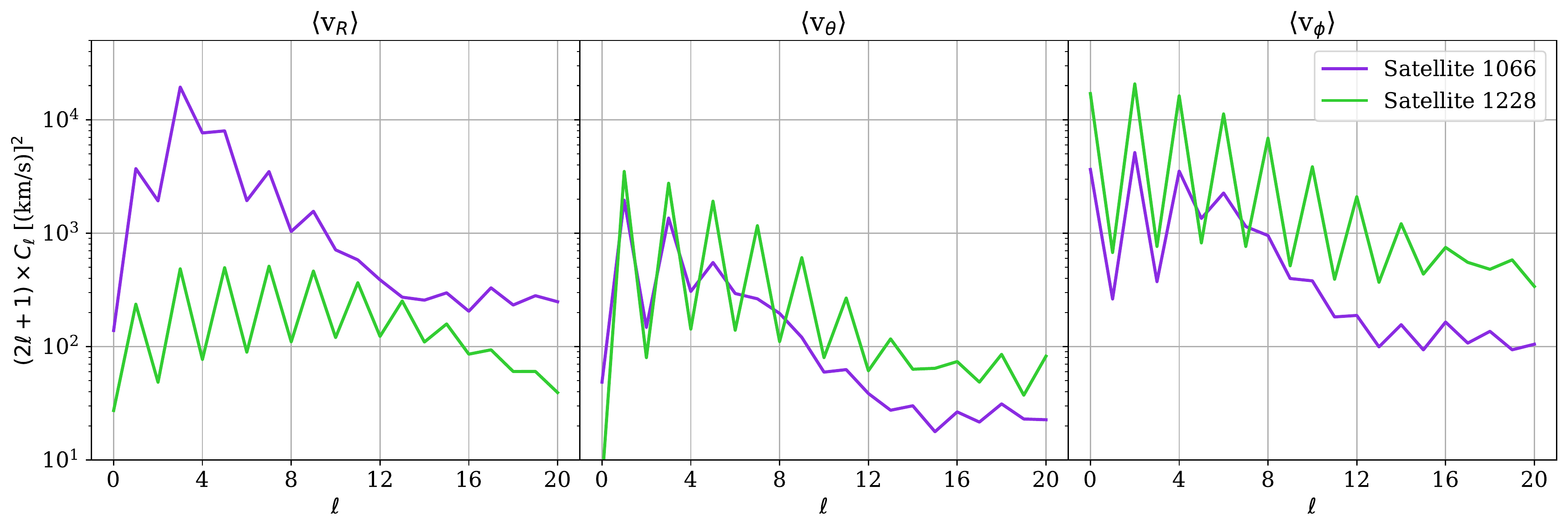}
    \caption{Power spectra for the accretion events mapped in Figure \ref{fig:bj_sat1066}. Both accretion events result in power spectra characterized by a sawtooth pattern, with peaks at odd $\ell$ in $v_R$ and $v_{\theta}$ and even $\ell$ in $v_{\phi}$.}
    \label{fig:bj05_ind_ps}
\end{figure*}

The power spectra for these velocity maps are shown as the purple curves are shown in Figure \ref{fig:bj05_ind_ps}. This radial accretion event has the most power in the radial velocity power spectrum; the power spectra peak at odd $\ell$ values for $v_R$ and $v_{\theta}$, while the $v_{\phi}$ power spectrum is dominated by peaks at even $\ell$ values.

As a different example, we can consider the debris from Satellite 1288, shown in the bottom panels of Figure \ref{fig:bj_sat1066}. This relatively low-mass ($M_{\rm Sat}=5.54 \times 10^{10}~M_{\odot}$), recently accreted ($t_{\rm Sat}=6.55~\rm{Gyr})$, circular $J_{\rm Sat}/J_{\rm circ}= 0.99$) accretion event leaves a w-wrapped kinematically cold stream that lies approximately in the $(x,y)$ plane. When we take the cross section of stars from $30-50$ kpc, we capture many of the stream stars, which form nearly continuous bands in the Mollweide projection. Because the stream is not perfectly circular, at opposite points in the orbit, the radial and polar velocities will be of similar magnitude but of opposite sign, while the tangential velocity is approximately constant along the stream. 

The power spectra are shown as the green curves in Figure \ref{fig:bj05_ind_ps}. This circular accretion event has more power in $v_{\theta}$ and $v_{\phi}$ than $v_{R}$. The $v_R$ and $v_{\theta}$ power spectra again shown peaks at odd $\ell$ values, while the $v_{\phi}$ power spectrum is peaked at odd $\ell$ values.

The fact that power spectra have these sawtooth patterns is indicative that spherical harmonic expansion of the different velocity components is not the best basis for Galactic substructure, given that the signatures in the angular power spectra are complicated. Potentially using vectorized spherical harmonics or a different coordinate transformation could provide ways forward to address this; we leave this to future work. 

\begin{figure}
    \centering
    \includegraphics[width=0.5\textwidth]{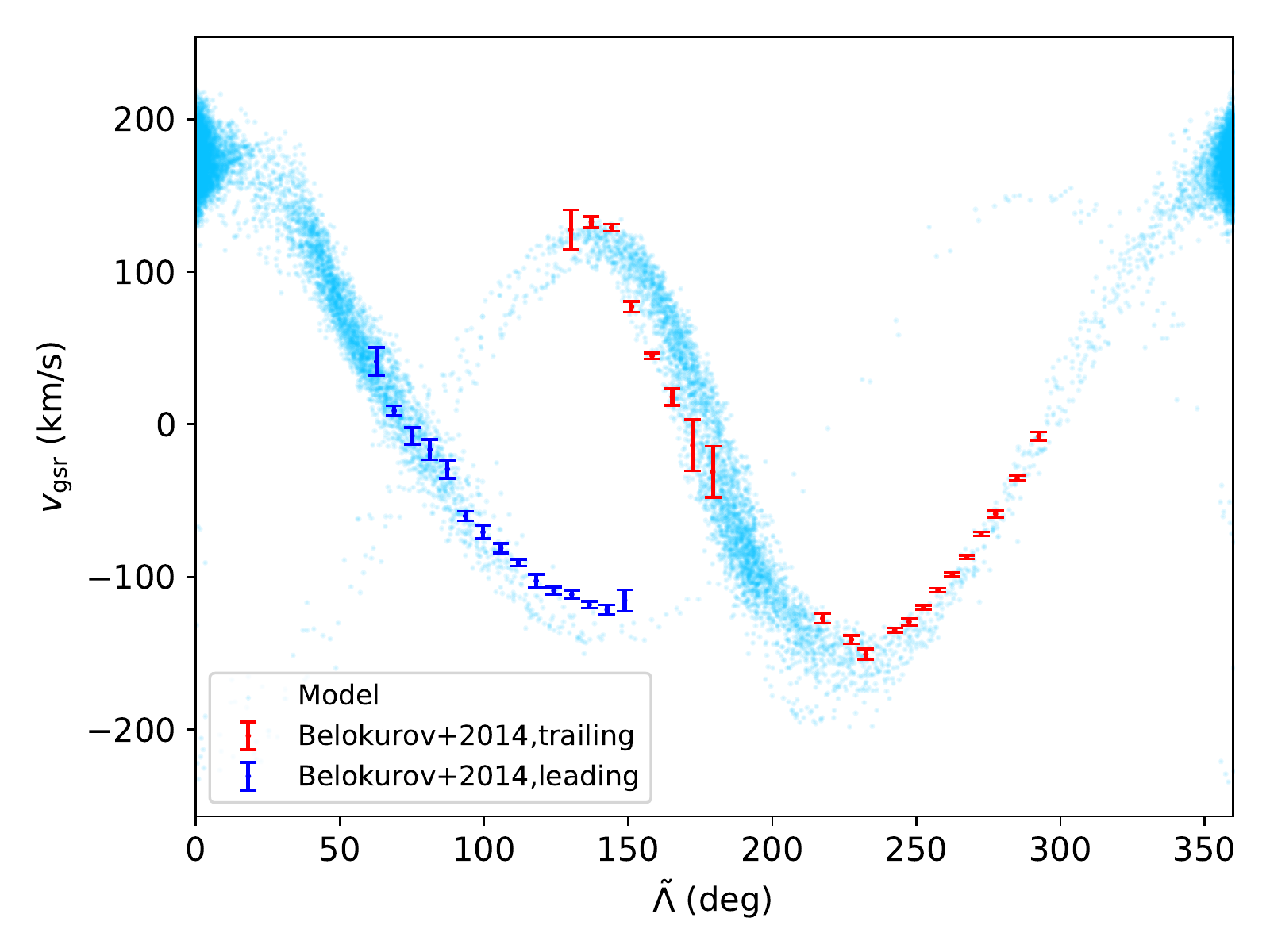}
    \caption{Comparison of best-fit Erkal model for Sgr with observed radial velocities in Sgr. We compare with radial velocities from \protect\cite{Belokurov2014} and use their coordinate system. We see that the model is a reasonable match to the data.}
    \label{fig:sgr_vgsr}
\end{figure}

\end{appendix}

\end{document}